\documentclass{article}

\usepackage{amsmath,amsfonts}
\usepackage{fullpage}
\usepackage{hyperref}
\usepackage{fancyhdr}
\DeclareMathOperator{\Tr}{Tr}
\DeclareMathOperator{\STr}{STr}
\newcommand{\phantomequals}{\mathrel{\hphantom{=}}}
\newcommand{\gYM}{g_{\text{YM}}}
\newcommand{\llangle}{\langle\hspace{-0.7mm}\langle}
\newcommand{\rrangle}{\rangle\hspace{-0.7mm}\rangle}

\title{Coupling M2-branes to background fields}
\author{James P Allen\footnote{Email: j.p.allen@durham.ac.uk}\; and Douglas J Smith\footnote{Email: douglas.smith@durham.ac.uk}}

\begin{document}

\maketitle

\thispagestyle{fancy}

\rhead{\large \bf DCPT-11/17}

\begin{center}

{\em Department of Mathematical Sciences,\\
Durham University,\\
South Road,\\
Durham.\\
DH1 3LE\\
UK}

\end{center}

\vspace{1.0cm}

\begin{abstract}
We discuss some of the issues arising in trying to extend the ABJM action to include couplings to background fields. This is analogous to the Myers-Chern-Simons terms of the multiple D$p$-brane action. We review and extend previous results to include terms which are quadratic in the background fields. These are fixed by requiring that we recover the correct Myers-type terms upon using the novel Higgs mechanism to reduce to the multiple D$2$-brane action.
\end{abstract}

\tableofcontents

\section{Introduction}

In the last two years, there has been much progress in understanding the worldvolume theory of multiple M$2$-branes. An important step in this was the formulation of explicit Lagrangian descriptions for the low energy dynamics of $N$ M$2$-branes. The first model proposed was by Bagger, Lambert and Gustavsson (BLG) based on the novel structure of a three-algebra \cite{Bagger:2007jr, Bagger:2007vi, Gustavsson:2007vu}. This has explicit $\mathcal{N} = 8$ supersymmetry and a gauge symmetry based on the three algebra. Unfortunately, it was shown that under some simple assumptions only one such three algebra exists and the theory only describes $2$ or $3$ coincident M$2$-branes \cite{Gauntlett:2008uf}.

Shortly after the introduction of the BLG theory, a full non-Abelian description of $N$ M$2$-branes was proposed by Aharony, Bergman, Jafferis and Maldacena (ABJM) \cite{Aharony:2008ug}. This is a $U(N) \times U(N)$ Chern-Simons gauge theory with levels $k$ and $-k$, and describes M$2$-branes sitting at the orbifold singularity of $\mathbb{C}^4 / \mathbb{Z}_k$. It only has $\mathcal{N} = 6$ manifest supersymmetry but this can be enhanced to $\mathcal{N} = 8$ using monopole operators when $k = 1,2$ \cite{Kwon:2009ar}.

It is possible to reduce the ABJM action to an $\mathcal{N} = 8$, $U(N)$ Yang-Mills gauge theory describing $N$ D$2$-branes by using the novel Higgsing mechanism originally introduced for the BLG action \cite{Mukhi:2008ux, Pang:2008hw}.

The full multiple D$2$-brane action includes couplings to the background fields of type IIA string theory. For a single brane this would be the pull back of $C_{(3)}$ to the world volume but for the non-Abelian multiple D$2$-brane action it must include further dielectric couplings to all of the R-R form fields, $C_{(1)}$, $C_{(5)}$, etc \cite{Myers:1999ps}. However, terms of this type are not present in either the BLG or ABJM actions and are not recovered by the Higgsing mechanism. M-theory contains a background $C_{(3)}$ form and its dual $C_{(6)}$ which should reduce to the R-R fields of string theory and it is interesting to ask how these might couple to the multiple M$2$-brane action in analogy to the D$2$-brane action. The form of the couplings to the BLG action has been explored in the presence of a constant field strength by preserving supersymmetry \cite{Lambert:2009qw}, and also by reducing to the D$2$-action \cite{Kim:2009nc}. 

In \cite{Kim:2010hj} the authors generalize the construction in \cite{Kim:2009nc} to the ABJM action. Their proposal for the form of a general background field coupling is guided by gauge invariance and the recovery of terms in the D$2$ action that are consistent with those already known. The correct couplings to $C_{(3)}$ and $C_{(5)}$ in the D$2$-brane action are recovered, as well as the coupling to the Kalb-Ramond field $B$. However, there are still some unresolved problems relating to how to write down gauge invariant couplings in the M$2$-brane action without over restricting the background form fields. So far it is only possible if the background fields are assumed to be at most linear functions in the coordinates.

All of the analysis so far has only considered couplings with a single background field term. However, the full dielectric coupling to the background fields in the multiple D$2$ action contains quadratic terms such as $C_{(3)} \wedge B$, and higher order terms. These should also be recoverable from the full M$2$-brane action. In this note we will extend the analysis of \cite{Kim:2010hj} to consider quadratic couplings to the background form fields and recover further terms of the multiple D$2$ action.

The next section is a brief review of the ABJM and its reduction to $\mathcal{N} = 8$, $U(N)$ Yang-Mills gauge theory via the novel Higgsing mechanism \cite{Mukhi:2008ux, Pang:2008hw}. We will also show how this is modified in the presence of couplings which are linear in the background fields \cite{Kim:2010hj}. The third and fourth sections show how we can extend the action to include quadratic couplings and fixed their coefficients by comparison with the expected terms in the D$2$ action.

\section{Review of ABJM theory coupled to background fields}

In this section we present a short review of the application of the novel Higgsing mechanism to ABJM's $\mathcal{N} = 6$, $U(N) \times U(N)$ Chern-Simons gauge theory. The procedure gives us an $\mathcal{N} = 8$, $U(N)$ Yang-Mills gauge theory and can be interpreted as recovering the world-volume theory of $N$ D$2$-branes from the world-volume theory of $N$ M$2$-branes. This was originally carried out in \cite{Pang:2008hw} and we paraphrase that work here using the notation and conventions established in \cite{Kim:2010hj}.

\subsection{The ABJM action}

The scalar fields in the ABJM action are combined into a $\mathbf{4}$ representation of $SU(4)$:
\begin{align}
Y^A = X^A + i X^{A+4}, \quad A = 1, \ldots, 4.
\end{align}
These are in the bifundamental representation $(N, \bar{N})$ of the $U(N) \times U(N)$ gauge group. We can split the $X^I$s into their trace and traceless parts:
\begin{align}
X^I &= \hat{x}^I + i \mathbf{x}^I, \qquad \qquad I = 1,\ldots, 8 \\
&= x^{0 I} T^0 + i x^{a I} T^a,
\end{align}
where 
\begin{gather}
T^0 = \text{diag}(1,\ldots,1), \\
\left[ T^a, T^b \right] = i f^{abc} T^c, \qquad \Tr(T^a) = 0,
\end{gather}
are the Hermitian generators of $U(N)$. Using these, we can separate out the Hermitian and anti-Hermitian parts of $Y^A$ and write $Y^A$ as
\begin{align}
Y^A &= \hat{x}^A + i \mathbf{x}^A + i \hat{x}^{A+4} - \mathbf{x}^{A+4} \\
&= \tilde X^A + i \tilde X^{A+4},
\end{align}
where
\begin{align}
\tilde X^A = \hat{x}^A - \mathbf{x}^{A+4}, \quad \text{and} \quad \tilde X^{A+4} = \hat{x}^{A + 4} + \mathbf{x}^{A},
\end{align}
are Hermitian fields.

There are two gauge potentials, $A^{(L)}_\mu$ and $A^{(R)}_\mu$ which transform in the adjoint of the left and right $U(N)$s respectively. The covariant derivative is given by
\begin{align}
D_\mu Y^A = \partial_\mu Y^A + i A^{(L)}_\mu Y^A - i Y^A A^{(R)}_\mu.
\end{align}

The action for the ABJM theory can be written as
\begin{align}
S = \int d^3x \left( - \Tr \left( D_\mu Y_A^\dagger D^\mu Y^A \right) +  i \Tr \left( \psi^{A \dagger} \gamma^\mu D_\mu \psi_A \right) - V_{\text{bos}} - V_{\text{ferm}} - \mathcal{L}_{\text{CS}} \right).
\end{align}
The gauge potentials have no degrees of freedom due to the presence of a Chern-Simons term. They come with equal but opposite sign Chern-Simons levels, $k$ and $-k$:
\begin{align}
\mathcal{L}_{\text{CS}} &= \frac{k}{4\pi} \epsilon^{\mu \nu \lambda} \Tr \left( A^{(L)}_\mu \partial_\nu A^{(L)}_\lambda + \frac{2i}{3} A^{(L)}_\mu A^{(L)}_\nu A^{(L)}_\lambda
- A^{(R)}_\mu \partial_\nu A^{(R)}_\lambda - \frac{2i}{3} A^{(R)}_\mu A^{(R)}_\nu A^{(R)}_\lambda \right).
\end{align}

The bosonic potential $V_{\text{bos}}$ is a sextic potential involving the scalars, and the fermionic potential $V_{\text{ferm}}$ contains Yukawa like terms mixing the scalars and fermions.

\subsection{Reduction to D$2$-branes}

There are two important steps in reducing the above action to the worldvolume theory of D$2$-branes. First we rewrite the gauge fields as
\begin{align}
A^{\pm}_\mu = \frac{1}{2} \left( A^{(L)}_\mu \pm A^{(R)}_\mu \right).
\end{align}
In this notation the covariant derivative becomes
\begin{align}
D_\mu Y^A &= \tilde D_\mu Y^A + i \left\{ A^{-}_\mu, Y^A \right\}, \quad \text{where} \quad \tilde D_\mu Y^A = \partial_\mu Y^A + i \left[ A^{+}_\mu, Y^A \right].
\end{align}
The Chern-Simons term becomes 
\begin{align}
\mathcal{L}_{\text{CS}} &= \frac{k}{2\pi} \epsilon^{\mu \nu \lambda} \Tr \left( A^{-}_\mu F^+_{\nu \lambda} + \frac{2}{3} i A^{-}_\mu A^{-}_\nu A^{-}_\lambda \right),
\end{align}
where 
\begin{align}
F^+_{\mu \nu} = \partial_\mu A^+_\nu - \partial_\nu A^+_\mu + i[A^+_\mu, A^+_\nu].
\end{align}

The other important step is to allow one of the scalar fields to acquire a large vacuum expectation value. In this case we take the VEV of $Y^4$ to be
\begin{align}
\langle Y^4 \rangle &= \frac{v}{2} T^0,
\end{align}
or equivalently,
\begin{align}
\langle \hat{x}^4 \rangle &= \frac{v}{2} T^0.
\end{align}
Doing so breaks the $U(N) \times U(N)$ gauge symmetry to its diagonal subgroup $U(N)$ where the left and right groups are identified.

The Yang-Mills coupling is defined by
\begin{align} \label{eq:g_{YM} definition}
\gYM &= \frac{2 \pi v}{k}.
\end{align}
Note that this differs from \cite{Pang:2008hw} but we could make the replacement $v \rightarrow \frac{2v}{\sqrt{2N}}$ to recover their normalizations and match their calculations in the rest of this note. We will however stick to the normalizations from \cite{Kim:2010hj}.

Following the novel Higgsing procedure, we have to take the limit $v \rightarrow \infty$ and $k \rightarrow \infty$ while keeping $\gYM$ fixed. Thus we will only keep the leading order terms in powers of $v^{-1}$ and $k^{-1}$.
When we integrate out $A^{-}_\mu$ using its equations of motion we find that $A^{-}_\mu$ is of order $v^{-1}$ and so the only terms left at leading order have equal powers of $A^{-1}_\mu$ and $v$ or $k$.
Bearing this in mind, and expanding around the expectation value $Y^4 \rightarrow Y^4 + \frac{v}{2} T^0$, the covariant derivatives become
\begin{align}
D_\mu Y^a &\rightarrow \tilde D_\mu \tilde X^a + i \tilde D_\mu \tilde X^{a + 4} + \mathcal{O} \left( v^{-1} \right), \quad a = 1, \ldots, 3 \\
D_\mu Y^4 &\rightarrow \tilde D_\mu \tilde X^4 + i \tilde D_\mu \tilde X^{8} + i v A^{-}_{\mu} + \mathcal{O} \left( v^{-1} \right).
\end{align}
We make the redefinition $A^{-}_\mu \rightarrow A^{-}_\mu - \frac{1}{v} \tilde D_\mu \tilde X^8$ so that the covariant derivatives are
\begin{align}
D_\mu Y^a &\rightarrow \tilde D_\mu \tilde X^a + i \tilde D_\mu \tilde X^{a + 4} + \mathcal{O} \left( v^{-1} \right), \quad a = 1, \ldots, 3 \label{eq:effective covariant derivative of Y^a}\\
D_\mu Y^4 &\rightarrow \tilde D_\mu \tilde X^4 + i v A^{-}_{\mu} + \mathcal{O} \left( v^{-1} \right). \label{eq:effective covariant derivative of Y^4}
\end{align}
This will also introduce a term into the action proportional to $\varepsilon^{\mu \nu \rho} \tilde D_\mu \tilde X^8 F^{+}_{\nu \rho}$ from the Chern-Simons term. However this can be written as a total derivative by use of partial integration and the Bianchi identity for $F^{+}_{\nu \rho}$.

Neglecting higher order terms in $v^{-1}$, the bosonic part of the action is thus
\begin{align}
S &= \int d^3x \Tr \left( - \tilde D_\mu \tilde X_i \tilde D^\mu \tilde X^i - v^2 A^{-}_\mu A^{- \mu} + \frac{k}{2 \pi} \varepsilon^{\mu \nu \lambda} A^{-}_\mu F^{+}_{\nu \lambda} \right) - V_{\text{bos}}.
\end{align}
We can eliminate $A^{-}_\mu$ from the action by solving for its equation of motion. Doing so, we find the equation of motion is
\begin{align}
A^{- \mu} &= \frac{k}{4 \pi v^2} \varepsilon^{\mu \nu \rho} F^{+}_{\nu \rho}.
\end{align}
Substituting this back into the action we have
\begin{align}
S &= \int d^3x \Tr \left( - \tilde D_\mu \tilde X_i \tilde D^\mu \tilde X^i - \frac{1}{2 \gYM^2} F^{+}_{\mu \nu} F^{+ \mu \nu} \right) - V_{\text{bos}},
\end{align}
where we have used the definition of $\gYM$ from \eqref{eq:g_{YM} definition}.
After rescaling $\tilde X \rightarrow \tilde X / \gYM$, the $SU(N)$ action becomes that of Yang-Mills gauge theory:
\begin{align}
S &= \int d^3x \frac{1}{\gYM^2} \Tr \left( - \tilde D_\mu \tilde X_i \tilde D^\mu \tilde X^i - \frac{1}{2} F^{+}_{\mu \nu} F^{+ \mu \nu} \right) - V_{\text{bos}}.
\end{align}

We have not shown them explicitly, but the potentials also reduce to the expected bosonic and fermionic potentials on the D$2$ worldvolume.

\subsection{Couplings to background form fields}

In \cite{Kim:2010hj}, the authors extended the ABJM action to include couplings to the $C_{(3)}$ and $C_{(6)}$ fields of M-theory. When this new action was reduced by following the same procedure as in the previous section, the additional terms were shown to reproduce the expected D$2$ brane couplings to the $C_{(3)}$ and $C_{(5)}$ R-R forms of type IIA string theory and to the Kalb-Ramond field, $B_{\mu \nu}$.

The world-volume theory of $N$ D-branes in the presence of the R-R and Kalb-Ramond background fields is well understood. A single D$p$ brane naturally couples to the $C_{p+1}$ form via the pull-back of the form to the brane worldvolume. The non-Abelian multiple D-brane action must also contain couplings to the other R-R fields in order to preserve T-duality \cite{Myers:1999ps}. These extra terms are constructed by contracting the background forms with $[X^i, X^j]$ to decrease the degree of the form, and taking the wedge product with $(F + B)$ to increase the degree of the form. Extending this analysis to the M$2$ brane is difficult since we will see that it is not clear how to consider the pull-back of $C_{(3)}$ and $C_{(6)}$ in the ABJM action.

In the Yang-Mills gauge theory of D$2$-branes there is an obvious interpretation of the scalars $\mathbf{x}^i$ as corresponding to the transverse directions in the extrinsic $10$ dimensional space-time. The scalars have an $SO(7)$ symmetry and together with the $SO(2,1)$ symmetry of the worldvolume coordinates $\mu, \nu, \rho = 0, \ldots, 2$ these form a subgroup of the Lorentz symmetry of the entire space time, $SO(2,1) \times SO(7) \subset SO(9,1)$. If we consider one of the background R-R forms of IIA string theory, say $C_{(3)}$, then we can write it with indices as $C_{IJK}$, where $I,J,K = 0,\ldots,9$. We can naturally split this up into $C_{\mu \nu \rho}$, $C_{\mu \nu i}$, $C_{\mu i j}$ and $C_{i j k}$, and write the pull back to the D$2$ world-volume as
\begin{align}
P\left[ C_{(3)} \right]_{\mu \nu \rho} = C_{\mu \nu \rho} + 3 C_{\mu \nu i} D_\rho \mathbf{x}^i + 3 C_{\mu i j} D_\nu \mathbf{x}^i D_\rho \mathbf{x}^j + C_{i j k} D_\mu \mathbf{x}^i D_\nu \mathbf{x}^j D_\rho \mathbf{x}^k.
\end{align}

However, in the ABJM action the equivalent scalar fields are in a representation of $SU(4)$. This does not have an obvious interpretation as a subgroup of the full space-time symmetry. It is not clear how to generalise pull-backs or how to construct a coupling to the $C_{(3)}$ and $C_{(6)}$ fields of M-theory. Since we cannot couple directly to the space-time indices of $C_{(3)}$ and $C_{(6)}$, we will instead consider fields that still have $3$ and $6$ indices, but with different combinations of the $SO(2,1)$ worldvolume indices $\mu,\nu,\rho$ and the $SU(4)$ R-symmetry indices $A, B, \bar{A}, \bar{B}$, for example, $C_{\mu \nu \rho}$, $C_{\mu \nu A}$, $C_{\mu A \bar{B}}$, etc.

The scalar fields $Y^A$ are in the bifundamental representation of $U(N) \times U(N)$ and to write a gauge invariant matrix product we must alternate bifundamental fields $Y^A$ with anti-bifundamental fields $Y^{\dagger \bar{A}}$. This restricts us to writing pull-back terms like
\begin{align} \label{eq:natural M2 pull-back}
C_{\mu A \bar{B}} D^\nu Y^A D^\rho Y^{\dagger \bar{B}}
\end{align}
inside the trace. In the reduction this will give us the $C_{\mu i j}$ and $B_{\mu i}$ terms in the D$2$ action, however there are not enough components of $C_{\mu A \bar{B}}$ to give independent components of $C_{\mu i j}$ and $B_{\mu i}$. With just this term, we would find that the $C_{\mu a b}$ components are related to the $C_{\mu a+4 b+4}$ components in the D$2$ action, where $a = 1, \ldots 3$. Clearly we need more fields in the M$2$ action.

After allowing $X^8$ to acquire a vacuum expectation value proportional to the identity matrix, the gauge symmetry is broken to a single $U(N)$. The left and right $U(N)$s are identified and the bifundamental fields are now in the adjoint of the remaining $U(N)$. At this point, we can allow terms like
\begin{align}
C_{\mu A B} D^\nu Y^A D^\rho Y^{B}
\end{align}
since this is gauge invariant under the single $U(N)$ symmetry. Together with \eqref{eq:natural M2 pull-back} this has enough components to produce all the fields in the D$2$ action.

Unfortunately it is not clear how to write down terms in the unbroken ABJM action which will reduce to give both of these terms after carrying out the first steps of the Higgsing procedure. A possible prescription is presented in \cite{Kim:2010hj} by allowing more general contractions between matrix indices rather than just matrix multiplication. However, this is unsatisfactory since the background fields are then restricted to be linear functions of the coordinates. Here we will only consider terms which we expect to appear after expanding around the expectation value and leave the full unbroken form of the M$2$ action as an open question.

The intermediate action for the $C_{(3)}$ fields obtained in \cite{Kim:2010hj} after gauge symmetry breaking has the following form (up to a differing factor of $3!$):
\begin{align} \label{eq:m2 linear C terms}
\begin{split}
S_{C} &= \mu_2 \int \varepsilon^{\mu \nu \rho} \Tr \Big(C_{\mu \nu \rho} + 3 \lambda C_{\mu \nu A} D_\rho Y^A  \\
& \hspace{1.2in} + 3 \lambda^2 \left( C^{(1)}_{\mu A \bar{B}} D_\nu Y^A D_\rho Y^{\dagger \bar{B}} - C^{(1)}_{\mu A \bar{B}} D_\nu Y^{\dagger \bar{B}} D_\rho Y^A + C^{(3)}_{\mu A B} D_\nu Y^A D_\rho Y^B \right) \\
& \hspace{1.2in} + \lambda^3 \big( C^{(1)}_{ABC} D_\mu Y^A D_\nu Y^B D_\rho Y^C + C^{(2)}_{AB\bar{C}} D_\mu Y^A D_\nu Y^B D_\rho Y^{\dagger \bar{C}} \\
& \hspace{1.6in} - C^{(2)}_{A C \bar{B}} D_\mu Y^A D_\nu Y^{\dagger \bar{B}} D_\rho Y^C + C^{(2)}_{B C \bar{A}} D_\mu Y^{\dagger \bar{A}} D_\nu Y^B D_\rho Y^C \big) \Big) + \text{(c.c.)},
\end{split}
\end{align}
where $\mu_2$ is the M$2$-brane tension, $\lambda = 2 \pi l_p^{3/2}$ and $l_p^{3/2}$ is the Planck length.. Here the $C$ terms with different index structures should be considered as different fields and the superscript numbers help to distinguish between fields with similar index structure. We use this notation to be consistent with \cite{Kim:2010hj}. The $C$ fields are anti-symmetric in any groups of identical types of indices. For example,
\begin{align}
C^{(3)}_{\mu A B} = C^{(3)}_{\mu [A B]}, \quad C^{(1)}_{ABC} = C^{(1)}_{[ABC]}, \quad C^{(2)}_{AB\bar{C}} = C^{(2)}_{[AB]\bar{C}}.
\end{align}

We perform the Higgsing procedure as above: expand around the expectation value; consider only leading order terms; and rescale $X \rightarrow X / \gYM$. After doing so and using equations \eqref{eq:effective covariant derivative of Y^a} and \eqref{eq:effective covariant derivative of Y^4} the action can be written as
\begin{align}
S_{C} &= \mu_2 \int \varepsilon^{\mu \nu \rho} \left( P[ \tilde C ]_{\mu \nu \rho} + \lambda v A^{-}_{\mu} P[ \tilde B ]_{\nu \rho} \right),
\end{align}
where the pull backs of the fields are given by,
\begin{align}
P[ \tilde C ]_{\mu \nu \rho} &= \left( \tilde C_{\mu \nu \rho} + 3 \tilde \lambda \tilde C_{\mu \nu i} \tilde D_\rho \tilde X^i + 3 \tilde \lambda^2 \tilde C_{\mu i j} \tilde D_\nu \tilde X^i \tilde D_\rho \tilde X^j
 + \tilde \lambda^3 \tilde C_{i j k} \tilde D_\mu \tilde X^i \tilde D_\nu \tilde X^j \tilde D_\rho \tilde X^k \right)
\end{align}
and
\begin{align}
P[ \tilde B ]_{\mu \nu} &= \left(\tilde B_{\mu\nu} + 2 \tilde \lambda \llangle \tilde B_{\mu i} \tilde D_\nu \tilde X^i \rrangle + \tilde \lambda^2 \llangle \tilde B_{ij} \tilde D_\nu \tilde X^i\tilde D_\rho \tilde X^j \rrangle \right).
\end{align}
We have defined
\begin{align}
\tilde \lambda = \frac{\lambda}{\gYM}, \quad \text{and} \quad \gYM = \frac{2 \pi v}{k},
\end{align}
and $\llangle \ldots \rrangle$ denotes the symmetric product.

The $\tilde C$ and $\tilde B$ fields are combinations of the fields in the original M$2$ action and are chosen to give the above action the correct form.\footnote{In general we will write the fields appearing in the D$2$ action with a tilde, and those appearing in the M$2$ action without.} Their expressions in terms of the original fields are written out fully in appendix \ref{identification of field components}.

When integrating out $A^{-}_{\mu}$ the equation of motion now becomes
\begin{align}
A^{- \mu} &= \frac{k}{4 \pi v^2} \varepsilon^{\mu \nu \rho} F^{+}_{\nu \rho} + \frac{\mu_2 \lambda}{2 v} \varepsilon^{\mu \nu \rho} P[ \tilde B_{\nu \rho} ]  \\
&= \frac{k}{4 \pi v^2} \varepsilon^{\mu \nu \rho} \left( F^{+}_{\nu \rho} + \mu_2 \tilde \lambda \gYM^2 P[ \tilde B_{\nu \rho} ] \right) \\
&= \frac{k}{4 \pi v^2} \varepsilon^{\mu \nu \rho} \left( F^{+}_{\nu \rho} + \frac{1}{\tilde \lambda} P[ \tilde B_{\nu \rho} ] \right),
\end{align}
where we have used
\begin{align}
\mu_2 = \frac{1}{\gYM^2 \tilde \lambda^2}.
\end{align}
Substituting back into the action gives
\begin{align}
\mathcal{L}_{SU(N)} &= \Tr \frac{1}{\gYM^2} \left( - \tilde D_\mu \tilde X_i \tilde D^\mu \tilde X^i
- \frac{1}{2} \left( F^{+}_{\mu \nu} + \frac{1}{\tilde \lambda} P[ \tilde B_{\mu \nu} ] \right)^2 \right) - V_{\text{bos}}.
\end{align}
This is the expected coupling to $\tilde B$ field in the D$2$ action, and we have seen above that we recovered the expected coupling to $\tilde C_{(3)}$ via its pull-back.

\section{Quadratic couplings between M2-branes and background fields}

The full expression for the couplings to the background R-R fields in the D$2$ action is given by the Myers-Chern-Simons term \cite{Myers:1999ps},
\begin{align} \label{eq:myers-chern-simons term}
\mu_2 \int \STr \left( P \left[ e^{\frac{i}{2} \tilde \lambda [X,X]} \left( \sum \tilde C_n e^{\tilde B} \right) \right] e^{\tilde \lambda \tilde F} \right).
\end{align}
Note that from now on we will drop the tilde on the $X$s for simplicity of notation, but we are referring to the Hermitian scalars that we have previously called $\tilde X$.
The M$2$ brane couplings considered so far only reproduce the single $C_{(3)}$ term in this. In \cite{Kim:2010hj} the authors also considered couplings to $C_{(6)}$ fields in the M$2$ action which reproduced the single $C_{(5)}$ term shown here. In this note we would like to extend the above analysis to consider quadratic couplings to the background fields. If we include terms in the M$2$ brane action with two $C_{(3)}$ terms, then we will be able to reproduce the $C \wedge B$ pieces that appear in \eqref{eq:myers-chern-simons term}.

Before we can construct possible quadratic terms we note that the fields in the D$2$ and M$2$ brane actions are matrix valued and so their order is important. In the Myers-Chern-Simons term of the D$2$-brane action  all combinations of $\tilde C$, $\tilde B$, $DX$ and $[X,X]$ are taken to be implicitly symmetrised, as denoted by the use of the \emph{symmetric} trace, $\STr$. Although we write terms in some canonical order we really mean their symmetric product:
\begin{align}
\tilde C \tilde B [X, X] \rightarrow \frac{1}{2} \left( \tilde C \tilde B [X, X] + \tilde B \tilde C [X, X] \right).
\end{align}
Cyclic permutations are taken into account by the trace which acts on all terms in the action.

We cannot naively impose the same prescription on the M$2$-brane action since taking two $C_{(3)}$ terms to be symmetrised leads to the term vanishing:
\begin{align}
C_{i j k} C_{l m o} &= - C_{l m o} C_{i j k}, &&\text{by anti-symmetry of indices} \\
&= - C_{i j k} C_{l m o}, &&\text{by symmetry of product}.
\end{align}
Instead we must take these products of $C$s to be anti-symmetric. This argument is perhaps too simple since we have a more complicated and mixed index structure on the $C$ fields than presented here. However, we will see later that this property is also required to match with the sign of certain terms in the reduction to the D$2$-brane action.

With this anti-symmetry property we only need to explicitly write half of the possible couplings in the M$2$-action. For example, we will only write $C_{\mu \nu \rho} C^{(1)}_{A B C}$ but this also includes $C^{(1)}_{A B C} C_{\mu \nu \rho}$ with the opposite sign.

\section{Reduction to D2-brane couplings}

The terms we would like to recover in the D$2$ action come from the $\tilde C \wedge \tilde B$ piece of \eqref{eq:myers-chern-simons term}:
\begin{align}
\tilde{\mathcal{L}} &= \frac{1}{2} \mu_2 \tilde \lambda i \Tr \left( P \left[ [X,X] \tilde C_3 \wedge \tilde B \right] \right) \\
&= \frac{\mu_2 \tilde \lambda}{2} i \varepsilon^{\mu \nu \rho} \Tr \left( \tilde C_{[\mu \nu \rho} \tilde B_{i j]} [X^i, X^j] + 3 \tilde \lambda \tilde C_{[\mu \nu i} \tilde B_{j k]} [X^i, X^j] \tilde D_\rho X^k + \ldots \right) \label{eq:pulled back MCS term}
\end{align}
In the following sections we will look at which terms we must add in to the M$2$ action to be able to recover each of the terms above. 
Nothing in the reduction procedure will remove or add derivatives so we can match the theories term by term based on the number of derivatives.

\subsection{No derivatives}

The piece of the D$2$ action that we are interested in with no derivatives is given by the first term of \eqref{eq:pulled back MCS term}. Expanding out the antisymmetry across the indices of $\tilde C$ and $\tilde B$ gives
\begin{align}
\frac{\mu_2 \tilde \lambda}{20} i \varepsilon^{\mu \nu \rho} \Tr \left[ \tilde C_{\mu \nu \rho} \tilde B_{i j} [X^i, X^j] - 6 \tilde C_{\mu \nu i} \tilde B_{\rho j} [X^i, X^j] + 3 \tilde C_{\mu i j} \tilde B_{\nu \rho} [X^i, X^j]\right]
\label{eq:zero derivatives MCS term}
\end{align}

To find the required terms in the M$2$ action which will reduce to the above expression, we can consider all possible quadratic couplings to the fields previously identified in \eqref{eq:m2 linear C terms} and fix their coefficients by reducing to the D$2$ action. Having already done this we will instead present the couplings with the correct coefficients and then show that these do indeed give the required parts of the D$2$ action. We propose the following terms for the no derivative part of the M$2$ action:
\begin{align}
\mathcal{L}_0 &= \mathcal{L}_1 + \mathcal{L}_2 + \mathcal{L}_3 + \mathcal{L}_4 + \text{(c.c.)},
\end{align}
where
\begin{align}
\mathcal{L}_1 &= - \frac{6}{5} \frac{\pi \mu_2 \lambda}{k} \varepsilon^{\mu \nu \rho} \Tr \left( C_{\mu \nu \rho} C^{(1)}_{A B C} \gamma^{ABC} - C_{\mu \nu \rho} C^{(2)}_{A B \bar{C}} \beta^{AB}_{C} \right), \\
\mathcal{L}_2 &= - \frac{6}{5} \frac{\pi \mu_2 \lambda}{k} \varepsilon^{\mu \nu \rho} \Tr \left( C_{\mu \nu \rho} C^{(1) \dagger}_{A B C} \gamma^{ABC \dagger} - C_{\mu \nu \rho} C^{(2) \dagger}_{A B \bar{C}} \beta^{AB \dagger}_{C} \right), \\
\mathcal{L}_3 &= \frac{36}{5} \frac{\pi \mu_2 \lambda}{k} \varepsilon^{\mu \nu \rho} \Tr \left( C_{\mu \nu A} C^{(1)}_{\rho B \bar{C}} \beta^{AB}_C - \frac{3}{2} C_{\mu \nu A} C^{(3)}_{\rho B C} \gamma^{ABC} \right), \label{eq:L_3 original} \\
\mathcal{L}_4 &= - \frac{36}{5} \frac{\pi \mu_2 \lambda}{k} \varepsilon^{\mu \nu \rho} \Tr \left( C_{\mu \nu A} C^{(1) \dagger}_{\rho C \bar{B}} \beta^{AB}_C - \frac{1}{2} C_{\mu \nu A} C^{(3) \dagger}_{\rho B C} \beta^{B C \dagger}_A \right), \label{eq:L4}
\end{align}
and
\begin{align}
\beta^{AB}_C &\equiv Y^{[A|} Y^\dagger_C Y^{|B]} = \frac{1}{2} \left( Y^A Y^\dagger_C Y^B - Y^B Y^\dagger_C Y^A \right), \label{eq:beta expression} \\
\gamma^{ABC} &\equiv Y^{[A} Y^{B} Y^{C]}. \label{eq:gamma expression}
\end{align}
In this expression the double $C$ terms are assumed to be anti-symmetrised. Note that the daggers also act on the indices so, for example, $C^{(1) \dagger}_{\rho C \bar{B}} = (C^{(1) \dagger})_{\rho \bar{C} B}$. We also use the notation $Y^{\dagger \bar{A}} = Y^\dagger_A$. We will see that $\mathcal{L}_1, \mathcal{L}_2$ and their conjugates give the first term in \eqref{eq:zero derivatives MCS term}, while $\mathcal{L}_3, \mathcal{L}_4$ and their conjugates will give the combination of the second and third term in \eqref{eq:zero derivatives MCS term}.

When we let $Y^4$ acquire a large vacuum expectation value, $\langle Y^4 \rangle = \frac{v}{2} T^0$, the only components of \eqref{eq:beta expression} and \eqref{eq:gamma expression} which will remain in the limit $v \rightarrow \infty$ are
\begin{align}
\beta^{a4}_4 &= \frac{v}{2} \left( [X^a, X^4] + i[X^{a + 4}, X^4] \right), \\
\beta^{ab}_4 &= \frac{v}{4} \left( [X^a, X^b] + i[X^{a + 4}, X^b] + i[X^{a}, X^{b + 4}] - [X^{a + 4}, X^{b + 4}] \right), \\
\beta^{a4}_b &= \frac{v}{4} \left( [X^a, X^b] + i[X^{a + 4}, X^b] - i[X^{a}, X^{b + 4}] + [X^{a + 4}, X^{b + 4}] \right), \\
\gamma^{ab4} &= \frac{v}{12} \left( [X^a, X^b] + i[X^{a + 4}, X^b] + i[X^{a}, X^{b + 4}] - [X^{a + 4}, X^{b + 4}] \right).
\end{align}
In this limit and after rescaling $X \rightarrow X / \gYM$, the expression $\mathcal{L}_1$ becomes
\begin{align}
\begin{split}
\mathcal{L}_1 &\rightarrow - \frac{6}{20} \frac{\pi \mu_2 \lambda v}{k g_{YM}^2} \varepsilon^{\mu \nu \rho} \Tr \Big( C_{\mu \nu \rho} \big( C^{(1)}_{ab4} \left( [X^a, X^b] + i[X^{a + 4}, X^b] + i[X^{a}, X^{b + 4}] - [X^{a + 4}, X^{b + 4}] \right) \\
&\hspace{1.5in} - C^{(2)}_{a b \bar{4}} \left( [X^a, X^b] + i[X^{a + 4}, X^b] + i[X^{a}, X^{b + 4}] - [X^{a + 4}, X^{b + 4}] \right) \\
&\hspace{1.5in} - 2 C^{(2)}_{a 4 \bar{b}} \left( [X^a, X^b] + i[X^{a + 4}, X^b] - i[X^{a}, X^{b + 4}] + [X^{a + 4}, X^{b + 4}] \right) \\
&\hspace{1.5in} - 4 C^{(2)}_{a 4 \bar{4}} \left( [X^a, X^4] + i[X^{a + 4}, X^4]\right) \big) \Big)
\end{split} \\
\begin{split}
&= - \tfrac{3}{20}  \mu_2 \tilde \lambda \varepsilon^{\mu \nu \rho} \Tr \Big( C_{\mu \nu \rho} \left( C^{(1)}_{ab4} - C^{(2)}_{a b \bar{4}} - C^{(2)}_{a 4 \bar{b}} + C^{(2)}_{b 4 \bar{a}} \right) [X^a, X^b] \\
&\hspace{1.2in} + 2 i C_{\mu \nu \rho} \left( C^{(1)}_{a b 4} - C^{(2)}_{a b \bar{4}} + C^{(2)}_{a 4 \bar{b}} + C^{(2)}_{b 4 \bar{a}} \right) [X^{a}, X^{b + 4}] \\
&\hspace{1.2in} - C_{\mu \nu \rho} \left( C^{(1)}_{a b 4} - C^{(2)}_{a b \bar{4}} + C^{(2)}_{a 4 \bar{b}} -  C^{(2)}_{b 4 \bar{a}}  \right) [X^{a + 4}, X^{b + 4}] \\
&\hspace{1.2in} - 4 C_{\mu \nu \rho} C^{(2)}_{a 4 \bar{4}} [X^a, X^4] - 4 i C_{\mu \nu \rho} C^{(2)}_{a 4 \bar{4}}  [X^{a + 4}, X^4] \Big).
\end{split}
\end{align}
We have made use of the anti-symmetry properties of $C^{(1)}$ and $C^{(2)}$ worked out in \cite{Kim:2010hj}.

Similarly, $\mathcal{L}_2$ becomes
\begin{align}
\begin{split}
\mathcal{L}_2 & \rightarrow \tfrac{3}{20} \mu_2 \tilde \lambda \varepsilon^{\mu \nu \rho} \Tr \Big( C_{\mu \nu \rho} \left( C^{(1) \dagger}_{ab4} - C^{(2) \dagger}_{a b \bar{4}} - C^{(2) \dagger}_{a 4 \bar{b}} + C^{(2) \dagger}_{b 4 \bar{a}} \right) [X^a, X^b] \\
&\hspace{1.2in} - 2 i C_{\mu \nu \rho} \left( C^{(1) \dagger}_{a b 4} - C^{(2) \dagger}_{a b \bar{4}} + C^{(2) \dagger}_{a 4 \bar{b}} + C^{(2) \dagger}_{b 4 \bar{a}} \right) [X^{a}, X^{b + 4}] \\
&\hspace{1.2in} - C_{\mu \nu \rho} \left( C^{(1) \dagger}_{a b 4} - C^{(2) \dagger}_{a b \bar{4}} + C^{(2) \dagger}_{a 4 \bar{b}} -  C^{(2) \dagger}_{b 4 \bar{a}} \right) [X^{a + 4}, X^{b + 4}] \\
&\hspace{1.2in} - 4 C_{\mu \nu \rho} C^{(2) \dagger}_{a 4 \bar{4}} [X^a, X^4] + 4 i C_{\mu \nu \rho} C^{(2) \dagger}_{a 4 \bar{4}}  [X^{a + 4}, X^4] \Big).
\end{split}
\end{align}

These terms are exactly the first term in equation \eqref{eq:zero derivatives MCS term} and thus give part of the no-derivative part of the pull back of $\tilde C \wedge \tilde B$ in the D$2$ action. To see this we split the $SO(7)$ indices, $i = 1,\ldots, 7$, into $a$, $4$ and $a+4$ where $a = 1,\ldots,3$. In this form we can substitute the $\tilde C$ and $\tilde B$ fields with their expressions in terms of the M2 $C$ fields. These have been previously determined by considering the reduction of the linear couplings and are given in Appendix \ref{identification of field components}. Thus the first term in equation \eqref{eq:zero derivatives MCS term} can be written as:
\begin{align}
&\phantomequals \frac{\mu_2 \tilde \lambda}{20} i \varepsilon^{\mu \nu \rho} \Tr \left( \tilde C_{\mu \nu \rho} \tilde B_{i j} [X^i, X^j] \right) \\
\begin{split}
&= \frac{\mu_2 \tilde \lambda}{20} i \varepsilon^{\mu \nu \rho} \Tr \Big( \tilde C_{\mu \nu \rho} \big( \tilde B_{ab} [X^a,X^b] + 2 \tilde B_{a4} [X^a, X^4] + 2 \tilde B_{a b+4} [X^a, X^{b+4}]\\
&\hspace{1.1in} - 2 \tilde B_{4 a+4} [X^{a + 4}, X^4] + \tilde B_{a + 4 b + 4} [X^{a + 4}, X^{b+4}] \big) \Big)
\end{split} \\
\begin{split}
&= - \tfrac{3}{20}  \mu_2 \tilde \lambda \varepsilon^{\mu \nu \rho} \Tr \Big( C_{\mu \nu \rho} \left( C^{(1)}_{ab4} - C^{(2)}_{a b \bar{4}} - C^{(2)}_{a 4 \bar{b}} + C^{(2)}_{b 4 \bar{a}} 
                                                                                                - C^{(1) \dagger}_{ab4} + C^{(2) \dagger}_{a b \bar{4}} + C^{(2) \dagger}_{a 4 \bar{b}} - C^{(2) \dagger}_{b 4 \bar{a}} \right) [X^a, X^b] \\
&\hspace{1.3in} + 2 i C_{\mu \nu \rho} \left( C^{(1)}_{ab4} - C^{(2)}_{a b \bar{4}} + C^{(2)}_{a 4 \bar{b}}  + C^{(2)}_{b 4 \bar{a}}
                                            + C^{(1) \dagger}_{ab4} - C^{(2) \dagger}_{a b \bar{4}} + C^{(2) \dagger}_{a 4 \bar{b}}  + C^{(2) \dagger}_{b 4 \bar{a}}\right) [X^{a}, X^{b + 4}] \\
&\hspace{1.3in} - C_{\mu \nu \rho} \left( C^{(1)}_{ab4} - C^{(2)}_{a b \bar{4}} + C^{(2)}_{a 4 \bar{b}} - C^{(2)}_{b 4 \bar{a}}
                                        - C^{(1) \dagger}_{ab4} + C^{(2) \dagger}_{a b \bar{4}} - C^{(2) \dagger}_{a 4 \bar{b}} + C^{(2) \dagger}_{b 4 \bar{a}} \right) [X^{a + 4}, X^{b + 4}] \\
&\hspace{1.3in} - 4 C_{\mu \nu \rho} \left(C^{(2)}_{a 4 \bar{4}} - C^{(2) \dagger}_{a 4 \bar{4}}\right) [X^a, X^4] - 4 i C_{\mu \nu \rho} \left( C^{(2)}_{a 4 \bar{4}} + C^{(2) \dagger}_{a 4 \bar{4}}\right) [X^{a + 4}, X^4].
\end{split}
\end{align}
This is indeed equal to $\mathcal{L}_1 + \mathcal{L}_2$. Only the undaggered terms in $\tilde C_{\mu \nu \rho}$ are considered here but the daggered terms will come from the conjugate of $\mathcal{L}_1 + \mathcal{L}_2$. Since both the M$2$ and D$2$ action are invariant under conjugation, these terms must also match in the same way.

Next we consider the terms in the proposed M$2$ action which mix the worldvolume and $SU(4)$ indices across $C$ terms:
\begin{align}
\mathcal{L}_3 &= \frac{\pi \mu_2 \lambda}{k} \frac{36}{5} \varepsilon^{\mu \nu \rho} \Tr \left( C_{\mu \nu A} C^{(1)}_{\rho B \bar{C}} \beta^{AB}_C - \tfrac{3}{2} C_{\mu \nu A} C^{(3)}_{\rho B C} \gamma^{ABC} \right) \\
\begin{split}
&\rightarrow \tfrac{9}{10} \mu_2 \tilde \lambda  \varepsilon^{\mu \nu \rho} \Tr \Big( C_{\mu \nu a} \left( C^{(1)}_{\rho 4 \bar{b}} + C^{(1)}_{\rho b \bar{4}} - C^{(3)}_{\rho b 4} \right) [X^a, X^b]
+ C_{\mu \nu a} \left( C^{(1)}_{\rho 4 \bar{b}} - C^{(1)}_{\rho b \bar{4}} + C^{(3)}_{\rho b 4} \right) [X^{a + 4}, X^{b + 4}] \\
&\hspace{1.1in} + i \Big( C_{\mu \nu a} C^{(1)}_{\rho b \bar{4}} - C_{\mu \nu b} C^{(1)}_{\rho a \bar{4}} - C_{\mu \nu a} C^{(1)}_{\rho 4 \bar{b}} - C_{\mu \nu b} C^{(1)}_{\rho 4 \bar{a}}
- C_{\mu \nu a} C^{(3)}_{\rho b 4} + C_{\mu \nu b} C^{(3)}_{\rho a 4} \Big) [X^a, X^{b + 4}] \\
&\hspace{1.1in} + 2 C_{\mu \nu a} C^{(1)}_{\rho 4 \bar{4}} [X^a, X^4] + 2i C_{\mu \nu a} C^{(1)}_{\rho 4 \bar{4}} [X^{a + 4}, X^4] \Big) \\
&\hspace{1.1in} - C_{\mu \nu 4} \left( C^{(1)}_{\rho a \bar{b}} + \tfrac{1}{2} C^{(3)}_{\rho a b} \right) [X^a, X^b] - i C_{\mu \nu 4} \left( - C^{(1)}_{\rho a \bar{b}} - C^{(1)}_{\rho b \bar{a}} + C^{(3)}_{\rho a b} \right) [X^a, X^{b + 4}] \\
&\hspace{1.1in} - C_{\mu \nu 4} \left( C^{(1)}_{\rho a \bar{b}} - \tfrac{1}{2} C^{(3)}_{\rho a b} \right) [X^{a + 4}, X^{b + 4}]  - 2 C_{\mu \nu 4} C^{(1)}_{\rho a \bar{4}} [X^a, X^4] \\
&\hspace{1.1in} - 2 i C_{\mu \nu 4} C^{(1)}_{\rho a \bar{4}} [X^{a + 4}, X^4] \Big).
\end{split}
\end{align}
This gives us the second and third terms of equation \eqref{eq:zero derivatives MCS term} and we see this by expanding and substituting as before:
\begin{align}
&\phantomequals - \frac{3}{10} \mu_2 \tilde \lambda i \varepsilon^{\mu \nu \rho} \Tr \left( \tilde C_{\mu \nu i} \tilde B_{\rho j} [X^i, X^j] \right) \\
\begin{split}
&=  - \tfrac{3}{10} \mu_2 \tilde \lambda i  \varepsilon^{\mu \nu \rho} \Tr \Big( \tilde C_{\mu \nu a} \tilde B_{\rho b} [X^a, X^b] + \left( \tilde C_{\mu \nu a} \tilde B_{\rho b+4} - \tilde C_{\mu \nu b+4} \tilde B_{\rho a} \right) [X^a, X^{b + 4}] \\
&\hspace{1.3in} + \tilde C_{\mu \nu a+4} \tilde B_{\rho b+4} [X^{a+4}, X^{b+4}] + \left( \tilde C_{\mu \nu a} \tilde B_{\rho 4} - \tilde C_{\mu \nu 4} \tilde B_{\rho a} \right) [X^a, X^4] \\
&\hspace{1.3in} + \left( \tilde C_{\mu \nu a+4} \tilde B_{\rho 4} - \tilde C_{\mu \nu 4} \tilde B_{\rho a+4} \right) [X^{a+4}, X^4] \Big)
\end{split} \\
\begin{split} \label{eq:expanded no-derivate second term}
&= \tfrac{9}{10} \mu_2 \tilde \lambda \varepsilon^{\mu \nu \rho} \Tr \Big( C_{\mu \nu a} \left( C^{(1)}_{\rho 4 \bar{b}} + C^{(1)}_{\rho b \bar{4}} - C^{(3)}_{\rho b 4} \right) [X^a, X^b]
+ C_{\mu \nu a} \left( C^{(1)}_{\rho 4 \bar{b}} - C^{(1)}_{\rho b \bar{4}} + C^{(3)}_{\rho b 4} \right) [X^{a+4}, X^{b+4}] \\
&\hspace{1.1in} - i \Big( C_{\mu \nu a} C^{(1)}_{\rho 4 \bar{b}} - C_{\mu \nu a} C^{(1)}_{\rho b \bar{4}} + C_{\mu \nu a} C^{(3)}_{\rho b 4}
+ C_{\mu \nu b} C^{(1)}_{\rho 4 \bar{a}} + C_{\mu \nu b} C^{(1)}_{\rho a \bar{4}} - C_{\mu \nu b} C^{(3)}_{\rho a 4} \Big) [X^a, X^{b + 4}] \\
&\hspace{1.1in} + 2 C_{\mu \nu a} C^{(1)}_{\rho 4 \bar{4}} [X^a, X^4] + 2 i C_{\mu \nu a} C^{(1)}_{\rho 4 \bar{4}} [X^{a + 4}, X^4] \\
&\hspace{1.1in} - C_{\mu \nu 4} \left( C^{(1)}_{\rho 4 \bar{a}} + C^{(1)}_{\rho a \bar{4}} - C^{(3)}_{\rho a 4} \right) [X^a, X^4]
- i C_{\mu \nu 4} \left( - C^{(1)}_{\rho 4 \bar{a}} + C^{(1)}_{\rho a \bar{4}} - C^{(3)}_{\rho a 4} \right) [X^{a + 4}, X^4] \Big)
\end{split}
\end{align}
and
\begin{align}
&\phantomequals\frac{3}{20} \mu_2 \tilde \lambda i  \varepsilon^{\mu \nu \rho} \Tr \left( \tilde B_{\mu \nu} \tilde C_{\rho i j} [X^i, X^j] \right) \\
\begin{split}
&= \frac{3}{20} \mu_2 \tilde \lambda i \varepsilon^{\mu \nu \rho} \Tr \Big( \tilde B_{\mu \nu} \tilde C_{\rho a b} [X^a, X^b] + 2 \tilde B_{\mu \nu} \tilde C_{\rho a b+4} [X^a, X^{b + 4}]  + \tilde B_{\mu \nu} \tilde C_{\rho a + 4 b + 4} [X^{a+4}, X^{b+4}] \\
&\hspace{1.2in} + 2 \tilde B_{\mu \nu} \tilde C_{\rho a 4} [X^a, X^4] + 2 \tilde B_{\mu \nu} \tilde C_{\rho 4 a+4} [X^4, X^{a+4}] \Big)
\end{split} \\
\begin{split} \label{eq:expanded no-derivate third term}
&= \frac{9}{10} \mu_2 \tilde \lambda  \varepsilon^{\mu \nu \rho} \Tr \Big( \tfrac{1}{2} C_{\mu \nu 4} \left( C^{(1)}_{\rho b \bar{a}} - C^{(1)}_{\rho a \bar{b}} - C^{(3)}_{\rho a b} \right) [X^a, X^b]
+ i C_{\mu \nu 4} \left( C^{(1)}_{\rho b \bar{a}} + C^{(1)}_{\rho a \bar{b}} - C^{(3)}_{\rho a b} \right) [X^a, X^{b + 4}] \\
&\hspace{1.2in} + \tfrac{1}{2} C_{\mu \nu 4} \left( C^{(1)}_{\rho b \bar{a}} - C^{(1)}_{\rho a \bar{b}} + C^{(3)}_{\rho a b} \right) [X^{a+4}, X^{b+4}]
- C_{\mu \nu 4} \left( - C^{(1)}_{\rho 4 \bar{a}} + C^{(1)}_{\rho a \bar{4}} + C^{(3)}_{\rho a 4} \right) [X^a, X^4] \\
&\hspace{1.2in} - i C_{\mu \nu 4} \left( C^{(1)}_{\rho 4 \bar{a}} + C^{(1)}_{\rho a \bar{4}} + C^{(3)}_{\rho a 4} \right) [X^{a + 4}, X^4] \Big). \\
\end{split}
\end{align}
The $C_{\mu \nu 4} \left(C^{(1)}_{\rho 4 \bar{a}} + C^{(3)}_{\rho a 4} \right) [X^{a + 4}, X^4]$ and $C_{\mu \nu 4} \left(C^{(1)}_{\rho 4 \bar{a}} - C^{(3)}_{\rho a 4} \right) [X^a, X^4]$ pieces cancel and do not appear in the overall action. This is remarkable as they are the only terms not recovered from possible couplings on the M$2$-brane. Every other term can be matched exactly with $\mathcal{L}_3$.
Note that we have only considered the undaggered terms here and will consider the daggered terms now.

The last terms from the no-derivative piece of the M$2$ action are given by $\mathcal{L}_4$ and contain one daggered $C$ and one undaggered $C$. We begin by looking at the second mixed term proposed in \eqref{eq:L4}:
\begin{align}
\frac{18}{5} \frac{\pi \mu_2 \lambda}{k} C_{\mu \nu A} C^{(3) \dagger}_{\rho B C} {\beta^{BC}_A}^\dagger.
\end{align}
In the reduction this gives the terms
\begin{align}
\begin{split}
&\frac{9}{10} \mu_2 \tilde \lambda \Big( C_{\mu \nu a} C^{(3) \dagger}_{\rho b 4} [X^a, X^b] + C_{\mu \nu a} C^{(3) \dagger}_{\rho b 4} [X^{a + 4}, X^{b + 4}]
- i \left( C_{\mu \nu a} C^{(3) \dagger}_{\rho b 4} + C_{\mu \nu b} C^{(3) \dagger}_{\rho a 4} \right) [X^a, X^{b+4}] \\
&\hspace{0.5in} - \tfrac{1}{2} C_{\mu \nu 4} C^{(3) \dagger}_{\rho a b} [X^a, X^b] + i C_{\mu \nu 4} C^{(3) \dagger}_{\rho a b} [X^a, X^{b + 4}] + \tfrac{1}{2} C_{\mu \nu 4} C^{(3) \dagger}_{\rho a b} [X^{a + 4}, X^{b + 4}] \\
&\hspace{0.5in} - 2 C_{\mu \nu 4} C^{(3) \dagger}_{\rho a 4} [X^a, X^4] + 2 i C_{\mu \nu 4} C^{(3) \dagger}_{\rho a 4} [X^{a + 4}, X^4] \Big).
\end{split}
\end{align}
While we have not written it explicitly, it is straight forward to see that these are exactly the terms that would appear in \eqref{eq:expanded no-derivate second term} and \eqref{eq:expanded no-derivate third term} if we included the $C^{(3) \dagger}$ from $\tilde C_{\mu i j}$ and $\tilde B_{\mu i}$.

The other mixed term is
\begin{align}
- \frac{36}{5} \frac{\pi \mu_2 \lambda}{k} C_{\mu \nu A} C^{(1) \dagger}_{\rho C \bar{B}} \beta^{AB}_C. \label{eq:daggered C1}
\end{align}
Whenever $C^{(1)}_{\mu a \bar{b}}$ appears in $\tilde C_{\mu i j}$ and $\tilde B_{\mu i}$ it always appears alongside its conjugate in the form $- C^{(1) \dagger}_{\mu b \bar{a}}$. Thus comparing this with the first term of \eqref{eq:L_3 original} we see that its inclusion in the M$2$ action will automatically give the correct $C^{(1) \dagger}$ terms in the D$2$ action.

The remaining terms in this part of the actions are given by conjugation of the existing ones. These must also match by the conjugation invariance of the actions.

\subsection{One derivative}

Conceptually we will follow the same procedure as before to identify the one derivative terms in the M$2$ action with the corresponding terms in the D$2$ action. However this quickly gets more complicated since we now have an extra $SU(4)$ or $SO(7)$ index in the product of $C$ fields. This will triple the number of ways we can split these terms into components with the indices, $a$, $4$, and $a + 4$. There are also now some terms where both $\tilde C$ fields are a sum of more than one of the original untilded fields. These gives many more cross terms when they are expanded out.

We first consider terms in the M$2$-action with the index structure $(\mu, \nu, A)$ and $(B, C, D)$:
\begin{align}
\begin{split}
& \frac{54}{5}  \frac{\pi \mu_2 \lambda^2}{k} \bigg[ \left( C_{\mu \nu A} C^{(1)}_{B C D} - \tfrac{1}{3} C_{\mu \nu D} C^{(1)}_{A B C} \right) \gamma^{ABC} D_{\rho} Y^D
+ C_{\mu \nu A} C^{(2)}_{B C \bar D} \gamma^{ABC} D_{\rho} Y^{\dagger \bar{D}} \\
& \hspace{0.7in} + \left( \tfrac{2}{3} C_{\mu \nu A} C^{(2)}_{B C \bar D} + \tfrac{1}{3} C_{\mu \nu C} C^{(2)}_{A B \bar D} \right) \beta^{AB}_{D} D_{\rho} Y^C \bigg].
\end{split}
\end{align}
In the reduction this becomes
\begin{align}
\begin{split}
&\tfrac{9}{10} \mu_2 \tilde \lambda^2 \Big(\left( \tfrac{1}{2} C_{\mu \nu 4} C^{(1)}_{a b c} - C_{\mu \nu a} C^{(1)}_{4 b c} - \tfrac{1}{2} C_{\mu \nu c} C^{(1)}_{a b 4} \right)
\left( [X^a, X^b] + i [X^{a+4}, X^b] + i [X^a, X^{b+4}] - [X^{a+4}, X^{b+4}]\right) \\
& \hspace{0.9in} \times \left( D_\rho X^c + i D_\rho X^{c+4} \right) \\
& \hspace{0.6in} + \left( C_{\mu \nu a} C^{(2)}_{b 4 \bar c} + \tfrac{1}{2} C_{\mu \nu 4} C^{(2)}_{a b \bar c} \right) \left( [X^a, X^b] + i [X^{a+4}, X^b] + i [X^a, X^{b+4}] - [X^{a+4}, X^{b+4}]\right) \\
& \hspace{1in} \times \left( D_\rho X^c - i D_\rho X^{c+4} \right) \\
& \hspace{0.6in} + \left(C_{\mu \nu a} C^{(2)}_{b c \bar 4} + \tfrac{1}{2} C_{\mu \nu c} C^{(2)}_{a b \bar 4} \right) \left( [X^a, X^b] + i [X^{a+4}, X^b] + i [X^a, X^{b+4}] - [X^{a+4}, X^{b+4}]\right) \\
& \hspace{1in} \times \left( D_\rho X^c + i D_\rho X^{c+4} \right) \\
& \hspace{0.6in} + \left(C_{\mu \nu a} C^{(2)}_{4 c \bar b} - C_{\mu \nu 4} C^{(2)}_{a c \bar b} + C_{\mu \nu c} C^{(2)}_{a 4 \bar b} \right) \left( [X^a, X^b] + i [X^{a+4}, X^b] - i [X^a, X^{b+4}] + [X^{a+4}, X^{b+4}]\right) \\
& \hspace{1in} \times \left( D_\rho X^c + i D_\rho X^{c+4} \right) \\
& \hspace{0.6in} + \left(C_{\mu \nu a} C^{(2)}_{4 c \bar 4} - C_{\mu \nu 4} C^{(2)}_{a c \bar 4} + C_{\mu \nu c} C^{(2)}_{a 4 \bar 4}\right) \left( [X^a, X^4] + i [X^{a+4}, X^4] \right) \left( D_\rho X^c + i D_\rho X^{c+4} \right)
\Big).
\end{split} \label{eq:m2 2-3 index piece}
\end{align}
Note that we have ignored the $D_\rho Y^4$ pieces for now.

The expected terms in the D$2$ action are
\begin{align}
&\tfrac{3}{2} \mu_2 \tilde \lambda^2 i \tilde C_{[\mu \nu i} \tilde B_{j k]} [X^i, X^j] \tilde D_\rho X^k \\
&= \tfrac{3}{20} \mu_2 \tilde \lambda^2 i \left( 2 \tilde C_{\mu \nu i} \tilde B_{j k} + \tilde C_{\mu \nu k} \tilde B_{i j}  + 2\tilde C_{\mu i j} \tilde B_{\nu k} - 4 \tilde C_{\mu i k} \tilde B_{\nu j} + \tilde C_{i j k} \tilde B_{\mu \nu} \right) [X^i, X^j] \tilde D_\rho X^k. \label{eq:d2 one derivative terms}
\end{align}
Looking at the index structure, the terms in \eqref{eq:m2 2-3 index piece} should equal the first, second and fifth term in \eqref{eq:d2 one derivative terms}. In the lists below, we list the index structure of the term we are considering and then its contribution when expanding out in terms of the original M$2$ fields. These can all then be seen to match with terms from the M$2$ action. We won't match every term but will present a selection to illustrate that the terms seem to all be matching. The first term from this part of the D$2$ action has contributions from
\begin{align}
\tilde C_{\mu \nu a} \tilde B_{b c} & \colon \quad
 - \tfrac{9}{10} \mu_2 \tilde \lambda^2 C_{\mu \nu a} \left(C^{(1)}_{b c 4} - C^{(2)}_{b c \bar 4} + C^{(2)}_{c 4 \bar b} - C^{(2)}_{b 4 \bar c} \right) [X^a, X^b] \tilde D_\rho X^c, \\
\tilde C_{\mu \nu a + 4} \tilde B_{b + 4 c} & \colon \quad
 \tfrac{9}{10} \mu_2 \tilde \lambda^2 C_{\mu \nu a} \left(C^{(1)}_{b c 4} - C^{(2)}_{b c \bar 4} - C^{(2)}_{c 4 \bar b} - C^{(2)}_{b 4 \bar c} \right) [X^{a+4}, X^{b+4}] \tilde D_\rho X^c, \\
\tilde C_{\mu \nu a} \tilde B_{b+4 c} & \colon \quad
 - \tfrac{9}{10} i \mu_2 \tilde \lambda^2 C_{\mu \nu a} \left(C^{(1)}_{b c 4} - C^{(2)}_{b c \bar 4} - C^{(2)}_{c 4 \bar b} - C^{(2)}_{b 4 \bar c} \right) [X^a, X^{b+4}] \tilde D_\rho X^c, \\
\tilde C_{\mu \nu a+4} \tilde B_{b c} & \colon \quad
 - \tfrac{9}{10} i \mu_2 \tilde \lambda^2 C_{\mu \nu a} \left(C^{(1)}_{b c 4} - C^{(2)}_{b c \bar 4} + C^{(2)}_{c 4 \bar b} - C^{(2)}_{b 4 \bar c} \right) [X^{a+4}, X^b] \tilde D_\rho X^c, \\
\tilde C_{\mu \nu a} \tilde B_{b c + 4} & \colon \quad
 - \tfrac{9}{10} i \mu_2 \tilde \lambda^2 C_{\mu \nu a} \left(C^{(1)}_{b c 4} - C^{(2)}_{b c \bar 4} + C^{(2)}_{c 4 \bar b} + C^{(2)}_{b 4 \bar c} \right) [X^a, X^b] \tilde D_\rho X^{c + 4}.
\end{align}
We have not written down $\tilde C_{\mu \nu a} \tilde B_{b + 4 c + 4}$,  $\tilde C_{\mu \nu a + 4} \tilde B_{b c + 4}$ or $\tilde C_{\mu \nu a+4} \tilde B_{b+4 c + 4}$ here.
The second term has contributions from
\begin{align}
\tilde C_{\mu \nu c} \tilde B_{a b} & \colon \quad
 - \tfrac{9}{20} \mu_2 \tilde \lambda^2 C_{\mu \nu c} \left(C^{(1)}_{a b 4} - C^{(2)}_{a b \bar 4} + C^{(2)}_{b 4 \bar a} - C^{(2)}_{a 4 \bar b} \right) [X^a, X^b] \tilde D_\rho X^c, \\
\tilde C_{\mu \nu c} \tilde B_{a b + 4} & \colon \quad
 - \tfrac{9}{20} i \mu_2 \tilde \lambda^2 C_{\mu \nu c} \left(C^{(1)}_{a b 4} - C^{(2)}_{a b \bar 4} + C^{(2)}_{b 4 \bar a} + C^{(2)}_{a 4 \bar b} \right) [X^a, X^{b + 4}] \tilde D_\rho X^c, \\
\tilde C_{\mu \nu c} \tilde B_{a+4 b} & \colon \quad
 - \tfrac{9}{20} i \mu_2 \tilde \lambda^2 C_{\mu \nu c} \left(C^{(1)}_{a b 4} - C^{(2)}_{a b \bar 4} - C^{(2)}_{b 4 \bar a} - C^{(2)}_{a 4 \bar b} \right) [X^{a+4}, X^b] \tilde D_\rho X^c, \\
\tilde C_{\mu \nu c + 4} \tilde B_{a + 4 b} & \colon \quad
 \tfrac{9}{20} \mu_2 \tilde \lambda^2 C_{\mu \nu c} \left(C^{(1)}_{a b 4} - C^{(2)}_{a b \bar 4} - C^{(2)}_{b 4 \bar a} - C^{(2)}_{a 4 \bar b} \right) [X^{a+4}, X^b] \tilde D_\rho X^{c + 4}.
\end{align}
The fifth term has contributions from
\begin{align}
\tilde C_{a b c} \tilde B_{\mu \nu} & \colon \quad
  - \tfrac{9}{20} \mu_2 \tilde \lambda^2 \left( C^{(1)}_{a b c} + C^{(2)}_{a b \bar c} - C^{(2)}_{a c \bar b} + C^{(2)}_{b c \bar a} \right) C_{\mu \nu 4} [X^a, X^b] \tilde D_{\rho} X^c, \\
\tilde C_{a+4 b c} \tilde B_{\mu \nu} & \colon \quad
  - \tfrac{9}{20} i \mu_2 \tilde \lambda^2 \left( C^{(1)}_{a b c} + C^{(2)}_{a b \bar c} - C^{(2)}_{b \bar c a} - C^{(2)}_{a \bar c b} \right) C_{\mu \nu 4} [X^{a+4}, X^b] \tilde D_{\rho} X^c, \\
\tilde C_{a b+4 c} \tilde B_{\mu \nu} & \colon \quad
  - \tfrac{9}{20} i \mu_2 \tilde \lambda^2 \left( C^{(1)}_{a b c} + C^{(2)}_{a b \bar c} + C^{(2)}_{b \bar c a} + C^{(2)}_{a \bar c b} \right) C_{\mu \nu 4} [X^a, X^{b+4}] \tilde D_{\rho} X^c, \\
\tilde C_{a+4 b+4 c} \tilde B_{\mu \nu} & \colon \quad
  \tfrac{9}{20} \mu_2 \tilde \lambda^2 \left( C^{(1)}_{a b c} + C^{(2)}_{a b \bar c} - C^{(2)}_{b \bar c a} + C^{(2)}_{a \bar c b} \right) C_{\mu \nu 4} [X^{a+4}, X^{b+4}] \tilde D_{\rho} X^c.
\end{align}
These can be identified with terms in the M$2$ action if we take into account the anti-symmetrised product on the $C$ fields.

In \eqref{eq:m2 2-3 index piece} we have ignored contributions from terms involving $D_\rho Y^4$ but we will demonstrate with a few examples that these also match. Considering the following terms which should also be included in the reduction \eqref{eq:d2 one derivative terms}:
\begin{align}
\begin{split}
&\tfrac{9}{10} \mu_2 \tilde \lambda^2 \left( 2 C_{\mu \nu a} C^{(2)}_{b 4 \bar 4} + C_{\mu \nu 4} C^{(2)}_{a b \bar 4} \right) \left( [X^a, X^b] + i [X^{a+4}, X^b] + i [X^a, X^{b+4}] - [X^{a+4}, X^{b+4}]\right) \tilde D_\rho X^4.
\end{split} \label{eq:m2 2-3 index piece DX^4 part}
\end{align}
If we expand out some more of the terms in \eqref{eq:d2 one derivative terms} we see that they do indeed match \eqref{eq:m2 2-3 index piece DX^4 part}. We have the following terms with only one $4$ index:
\begin{align}
\tilde C_{\mu \nu a} \tilde B_{b 4} & \colon \quad
  \tfrac{9}{5} \mu_2 \tilde \lambda^2 C_{\mu \nu a} C^{(2)}_{b 4 \bar 4} [X^a, X^b] \tilde D_{\rho} X^4, \\
\tilde C_{\mu \nu a} \tilde B_{b+4 4} & \colon \quad
  \tfrac{9}{5} \mu_2 \tilde \lambda^2 i C_{\mu \nu a} C^{(2)}_{b 4 \bar 4} [X^a, X^{b+4}] \tilde D_{\rho} X^4, \\
\tilde C_{\mu \nu a+4} \tilde B_{b 4} & \colon \quad
  \tfrac{9}{5} \mu_2 \tilde \lambda^2 i C_{\mu \nu a} C^{(2)}_{b 4 \bar 4} [X^{a+4}, X^b] \tilde D_{\rho} X^4, \\
\tilde C_{\mu \nu a+4} \tilde B_{b+4 4} & \colon \quad
  - \tfrac{9}{5} \mu_2 \tilde \lambda^2 C_{\mu \nu a} C^{(2)}_{b 4 \bar 4} [X^{a+4}, X^{b+4}] \tilde D_{\rho} X^4,
\end{align}
which are all included in \eqref{eq:m2 2-3 index piece DX^4 part}.
From the second and fifth term in \eqref{eq:d2 one derivative terms} we have some fortunate cancellations to give us more terms appearing in \eqref{eq:m2 2-3 index piece DX^4 part}:
\begin{align}
\begin{split}
\tilde C_{\mu \nu 4} \tilde B_{a b} + \tilde C_{a b 4} \tilde B_{\mu \nu} \colon \quad &
  - \tfrac{9}{20} \mu_2 \tilde \lambda^2 C_{\mu \nu 4} \left( C^{(1)}_{a b 4} - C^{(2)}_{a b \bar 4} + C^{(2)}_{b \bar 4 a} - C^{(2)}_{a \bar 4 b} \right) [X^a, X^b] \tilde D_{\rho} X^4 \\
& \qquad - \tfrac{9}{20} \mu_2 \tilde \lambda^2 \left( C^{(1)}_{a b 4} + C^{(2)}_{a b \bar 4} + C^{(2)}_{b \bar 4 a} - C^{(2)}_{a \bar 4 b} \right) C_{\mu \nu 4} [X^a, X^b] \tilde D_{\rho} X^4 \\
&= \tfrac{9}{10} \mu_2 \tilde \lambda^2 C_{\mu \nu 4}  C^{(2)}_{a b \bar 4} [X^a, X^b] \tilde D_{\rho} X^4.
\end{split}
\end{align}
There are no $C_{\mu \nu 4} C^{(2)}_{i 4 \bar 4}$ terms remaining in \eqref{eq:m2 2-3 index piece DX^4 part} and indeed we see that the terms from the D$2$-action that would contribute such terms will cancel with each other:
\begin{align}
\left( 2 \tilde C_{\mu \nu i} \tilde B_{j k} + \tilde C_{\mu \nu k} \tilde B_{i j} \right) [X^i, X^j] \tilde D_\rho X^k
&\rightarrow \left( - 2 \tilde C_{\mu \nu 4} \tilde B_{i 4} + 2 \tilde C_{\mu \nu 4} \tilde B_{i 4} \right) [X^i, X^4] \tilde D_\rho X^4 \\
& = 0.
\end{align}

Thus it appears that the first, second and fifth term of \eqref{eq:d2 one derivative terms} are recovered from our proposed M$2$ couplings.

The third and fourth terms in the D$2$-action \eqref{eq:d2 one derivative terms} are more complicated because both the $\tilde C$ and $\tilde B$ fields are a sum of multiple fields from the M$2$ action and their expansion introduces many cross terms. Considering only a few of the total components, these terms give us the following expression (note that the daggered terms (d.t.) are either identical or come with a sign change compared to their equivalent undaggered terms):
\begin{align} \label{eq:d2 1 derivative mixed indices}
&\phantomequals \tfrac{3}{20} \mu_2 \tilde \lambda^2 i \left(  2\tilde C_{\mu i j} \tilde B_{\mu k} - 4 \tilde C_{\mu i k} \tilde B_{\nu j} \right) [X^i, X^j] \tilde D_\rho X^k \\
& \rightarrow \tfrac{3}{10} \mu_2 \tilde \lambda^2 i \left( \tilde C_{\mu a b} \tilde B_{\mu c} - 2 \tilde C_{\mu a c} \tilde B_{\nu b} \right) [X^a, X^b] \tilde D_\rho X^c + \cdots \\
\begin{split}
&= - \tfrac{9}{10} \mu_2 \tilde \lambda^2 \bigg( \Big( \left( C^{(1)}_{\mu a \bar b} - C^{(1)}_{\mu b \bar a} + C^{(3)}_{\mu a b} + \text{(d.t.)} \right) \left( C^{(1)}_{\nu 4 \bar c} + C^{(1)}_{\nu c \bar 4} - C^{(3)}_{\nu c 4} - \text{(d.t.)} \right) \\
&\hspace{1in} -2 \left( C^{(1)}_{\mu a \bar c} - C^{(1)}_{\mu c \bar a} + C^{(3)}_{\mu a c} + \text{(d.t.)} \right) \left( C^{(1)}_{\nu 4 \bar b} + C^{(1)}_{\nu b \bar 4} - C^{(3)}_{\nu b 4} - \text{(d.t.)} \right) \Big) [X^a, X^b] \tilde D_\rho X^c \\
&\hspace{0.8in} + \Big( \left( - C^{(1)}_{\mu a \bar b} - C^{(1)}_{\mu b \bar a} + C^{(3)}_{\mu a b} - \text{(d.t.)} \right) \left( C^{(1)}_{\nu 4 \bar c} + C^{(1)}_{\nu c \bar 4} - C^{(3)}_{\nu c 4} - \text{(d.t.)} \right) \\
&\hspace{1in} + 2 \left( C^{(1)}_{\mu a \bar c} - C^{(1)}_{\mu c \bar a} + C^{(3)}_{\mu a c} + \text{(d.t.)} \right) \left( C^{(1)}_{\nu 4 \bar b} - C^{(1)}_{\nu b \bar 4} + C^{(3)}_{\nu b 4} + \text{(d.t.)} \right) \Big) i [X^a, X^{b+4}] \tilde D_\rho X^c \\
&\hspace{0.8in} + \Big( \left( C^{(1)}_{\mu a \bar b} + C^{(1)}_{\mu b \bar a} - C^{(3)}_{\mu a b} - \text{(d.t.)} \right) \left( C^{(1)}_{\nu 4 \bar c} + C^{(1)}_{\nu c \bar 4} - C^{(3)}_{\nu c 4} - \text{(d.t.)} \right) \\
&\hspace{1in} - 2 \left( C^{(1)}_{\mu a \bar c} + C^{(1)}_{\mu c \bar a} + C^{(3)}_{\mu a c} - \text{(d.t.)} \right) \left( C^{(1)}_{\mu 4 \bar b} + C^{(1)}_{\mu b \bar 4} - C^{(3)}_{\mu b 4} - \text{(d.t.)} \right) i [X^{a+4}, X^b] \tilde D_\rho X^c \\
&\hspace{0.8in} + \cdots
\end{split} \\
\begin{split}
&= - \tfrac{9}{10} \mu_2 \tilde \lambda^2 \bigg( \Big( \left( C^{(1)}_{\mu a \bar b} - C^{(1)}_{\mu b \bar a} + C^{(3)}_{\mu a b} + \text{(d.t.)} \right) \left( C^{(1)}_{\nu 4 \bar c} + C^{(1)}_{\nu c \bar 4} - C^{(3)}_{\nu c 4} - \text{(d.t.)} \right) \\
&\hspace{1in} -2 \left( C^{(1)}_{\mu a \bar c} - C^{(1)}_{\mu c \bar a} + C^{(3)}_{\mu a c} + \text{(d.t.)} \right) \left( C^{(1)}_{\nu 4 \bar b} + C^{(1)}_{\nu b \bar 4} - C^{(3)}_{\nu b 4} - \text{(d.t.)} \right) \Big) [X^a, X^b] \tilde D_\rho X^c \\
&\hspace{0.8in} + \Big( 2 \left( - C^{(1)}_{\mu a \bar b} - C^{(1)}_{\mu b \bar a} + C^{(3)}_{\mu a b} - \text{(d.t.)} \right) \left( C^{(1)}_{\nu 4 \bar c} + C^{(1)}_{\nu c \bar 4} - C^{(3)}_{\nu c 4} - \text{(d.t.)} \right) \\
&\hspace{1in} + 2 \left( C^{(1)}_{\mu a \bar c} - C^{(1)}_{\mu c \bar a} + C^{(3)}_{\mu a c} + \text{(d.t.)} \right) \left( C^{(1)}_{\nu 4 \bar b} - C^{(1)}_{\nu b \bar 4} + C^{(3)}_{\nu b 4} + \text{(d.t.)} \right) \\
&\hspace{1in} + 2 \left( C^{(1)}_{\mu b \bar c} + C^{(1)}_{\mu c \bar b} + C^{(3)}_{\mu b c} - \text{(d.t.)} \right) \left( C^{(1)}_{\mu 4 \bar a} + C^{(1)}_{\mu a \bar 4} - C^{(3)}_{\mu a 4} - \text{(d.t.)} \right) \Big) i [X^a, X^{b+4}] \tilde D_\rho X^c \\
&\hspace{0.8in} + \cdots
\end{split}
\end{align}

The pieces mixing $C^{(1)}$ and $C^{(3)}$ can be recovered from the following M$2$ action couplings:
\begin{align}
& \phantomequals - \frac{108}{5} \frac{\pi \mu_2 \lambda^2}{k} C^{(3)}_{\mu [A B |}C^{(1)}_{\nu | C ] \bar D} \beta^{BC}_D D_\rho Y^A \\
& = - \frac{108}{5} \frac{\pi \mu_2 \lambda^2}{k} \left( \tfrac{2}{3} C^{(3)}_{\mu A B}C^{(1)}_{\nu C \bar D} \beta^{BC}_D D_\rho Y^A + \tfrac{1}{3} C^{(3)}_{\mu B C}C^{(1)}_{\nu A \bar D} \beta^{BC}_D D_\rho Y^A \right) \\
\begin{split}
& \rightarrow - \tfrac{9}{5} \mu_2 \tilde \lambda^2 \bigg( \left( - C^{(1)}_{\mu a \bar b} C^{(3)}_{\nu c 4}  - C^{(3)}_{\mu a c} C^{(1)}_{\nu 4 \bar b} + C^{(1)}_{\mu c \bar b} C^{(3)}_{\nu a 4} \right) \left( [X^a, X^b] + i [X^{a+4}, X^b] - i [X^a, X^{b+4}] + [X^{a+4}, X^{b+4}] \right) \\
& \hspace{1in} \times \left(\tilde D_\rho X^c + i \tilde D_\rho X^{c + 4} \right) \\
& \hspace{0.8in} + \left( - C^{(3)}_{\mu a c} C^{(1)}_{\nu b \bar 4} + \tfrac{1}{2} C^{(3)}_{\mu a b} C^{(1)}_{\nu c \bar 4} \right) \left( [X^a, X^b] + i [X^{a+4}, X^b] + i [X^a, X^{b+4}] - [X^{a+4}, X^{b+4}] \right) \\
& \hspace{1in} \times \left(\tilde D_\rho X^c + i \tilde D_\rho X^{c + 4} \right) \\
& \hspace{0.8in} + 2 \left( - C^{(3)}_{\mu c 4} C^{(1)}_{\nu a \bar 4} - C^{(3)}_{\mu a c} C^{(1)}_{\nu 4 \bar 4} + C^{(3)}_{\mu a 4} C^{(1)}_{\nu c \bar 4} \right) \left( [X^a, X^4] + i [X^{a+4}, X^4] \right) \left(\tilde D_\rho X^c + i \tilde D_\rho X^{c + 4} \right) \bigg),
\end{split}
\end{align}
and
\begin{align}
& \phantomequals - \frac{108}{5} \frac{\pi \mu_2 \lambda^2}{k} C^{(3)}_{\mu A B} C^{(1)}_{\nu C \bar D} \gamma^{ABC} D_\rho Y^\dagger_D \\
\begin{split}
& \rightarrow - \tfrac{9}{10} \mu_2 \tilde \lambda^2 \bigg( \left( - 2 C^{(1)}_{\mu b \bar c} C^{(3)}_{\nu a 4} + C^{(3)}_{\mu a b} C^{(1)}_{\nu 4 \bar c} \right) \left( [X^a, X^b] + i [X^{a+4}, X^b] + i [X^a, X^{b+4}] - [X^{a+4}, X^{b+4}] \right) \\
& \hspace{1in} \times \left(\tilde D_\rho X^c - i \tilde D_\rho X^{c + 4} \right).
\end{split}
\end{align}
Again these can be identified assuming the $C$ field product is anti-symmetrised.

The quadratic $C^{(1)}$ pieces can be recovered from the following terms in the M$2$ action:
\begin{align}
&\phantomequals - \frac{72}{5} \frac{\pi \mu_2 \lambda^2}{k} \left( C^{(1)}_{\mu A \bar B} C^{(1)}_{\nu C \bar D} \beta^{A C}_B D_\rho Y^{\dagger}_D + C^{(1)}_{\mu A \bar B} C^{(1)}_{\nu C \bar D} \beta^{BD \dagger}_A D_\rho Y^C \right) \\
\begin{split}
&= - \frac{72}{5} \frac{\pi \mu_2 \lambda^2}{k} \Big( \left( C^{(1)}_{\mu a \bar b} C^{(1)}_{\nu 4 \bar c} - C^{(1)}_{\mu 4 \bar b} C^{(1)}_{\nu a \bar c} \right) \beta^{a 4}_b D_\rho Y^{\dagger}_c + C^{(1)}_{\mu a \bar 4} C^{(1)}_{\nu b \bar c} \beta^{a b}_4 D_\rho Y^{\dagger}_c \\
&\hspace{1.2in} + \left( C^{(1)}_{\mu b \bar a} C^{(1)}_{\nu c \bar 4} - C^{(1)}_{\mu b \bar 4} C^{(1)}_{\nu c \bar a} \right) \beta^{a 4 \dagger}_b D_\rho Y^c + C^{(1)}_{\mu 4 \bar a} C^{(1)}_{\nu c \bar b} \beta^{ab \dagger}_4 D_\rho Y^c + \cdots \Big)
\end{split}  \\
\begin{split}
&\rightarrow - \tfrac{9}{5} \mu_2 \tilde \lambda^2  \Big( \left( C^{(1)}_{\mu a \bar b} C^{(1)}_{\nu 4 \bar c} - C^{(1)}_{\mu a \bar c} C^{(1)}_{\nu 4 \bar b}  \right) \left( [X^a, X^b] + i [X^{a+4}, X^b] - i [X^a, X^{b+4}] + [X^{a+4}, X^{b+4}] \right) \\
& \hspace{1.2in} \times \left(\tilde D_\rho X^c - i \tilde D_\rho X^{c + 4} \right) \\
& \hspace{0.8in} + C^{(1)}_{\mu b \bar c} C^{(1)}_{\nu a \bar 4} \left( [X^a, X^b] + i [X^{a+4}, X^b] + i [X^a, X^{b+4}] - [X^{a+4}, X^{b+4}] \right) \\
& \hspace{1.2in} \times \left(\tilde D_\rho X^c - i \tilde D_\rho X^{c + 4} \right) \\
& \hspace{0.8in}  - \left( C^{(1)}_{\mu b \bar a} C^{(1)}_{\nu c \bar 4} - C^{(1)}_{\mu c \bar a} C^{(1)}_{\nu b \bar 4} \right) \left([X^a, X^b] - i [X^{a+4}, X^b] + i [X^a, X^{b+4}] + [X^{a+4}, X^{b+4}] \right) \\
& \hspace{1.2in} \times \left(\tilde D_\rho X^c + i \tilde D_\rho X^{c + 4} \right) \\
& \hspace{0.8in}  - C^{(1)}_{\mu c \bar b} C^{(1)}_{\nu 4 \bar a} \left([X^a, X^b] - i [X^{a+4}, X^b] - i [X^a, X^{b+4}] - [X^{a+4}, X^{b+4}] \right) \\
& \hspace{1.2in} \times \left(\tilde D_\rho X^c + i \tilde D_\rho X^{c + 4} \right).
\end{split}
\end{align}
Also the quadratic $C^{(3)}$ pieces can be recovered from the following terms in the M$2$ action:
\begin{align}
&\phantomequals \frac{108}{5} \frac{\pi \mu_2 \lambda^2}{k} C^{(3)}_{\mu A \bar B} C^{(3)}_{\nu C \bar D} \gamma^{A B C} D_\rho Y^D \\
&\rightarrow - \frac{108}{5} \frac{\pi \mu_2 \lambda^2}{k} \left( C^{(3)}_{\mu a b} C^{(3)}_{\nu c \bar 4} + 2 C^{(3)}_{\mu a 4} C^{(3)}_{\nu b c} \right) \gamma^{ab4} D_\rho Y^c \\
\begin{split}
&\rightarrow - \tfrac{9}{10} \mu_2 \tilde \lambda^2  \left( C^{(3)}_{\mu a b} C^{(3)}_{\nu c \bar 4} + 2  C^{(3)}_{\mu b c} C^{(3)}_{\nu a 4} \right) \left( [X^a, X^b] + i [X^{a+4}, X^b] + i [X^a, X^{b+4}] - [X^{a+4}, X^{b+4}] \right) \\
& \hspace{1.2in} \times \left(\tilde D_\rho X^c + i \tilde D_\rho X^{c + 4} \right).
\end{split}
\end{align}

To obtain the mixed daggered pieces we consider the following terms in the M$2$-action:
\begin{align}
&\phantomequals \frac{108}{5} \frac{\pi \mu_2 \lambda^2}{k} C^{(3) \dagger}_{\mu A B} C^{(1)}_{\nu C \bar D} \gamma^{ABD \dagger} D_\rho Y^C \\
& = \frac{108}{5} \frac{\pi \mu_2 \lambda^2}{k} \left( -2 C^{(3) \dagger}_{\mu a 4} C^{(1)}_{\nu c \bar b} + C^{(3) \dagger}_{\mu a b} C^{(1)}_{\nu c \bar 4} \right) \gamma^{a b 4 \dagger} D_\rho Y^c \\
\begin{split}
& \rightarrow \tfrac{9}{10} \mu_2 \tilde \lambda^2 \left( 2 C^{(1)}_{\mu c \bar b} C^{(3) \dagger}_{\nu a 4} - C^{(3) \dagger}_{\mu a b} C^{(1)}_{\nu c \bar 4} \right) \left( [X^a, X^b] - i[X^{a+4}, X^b] - i[X^a, X^{b+4}] - [X^{a+4}, X^{b+4}] \right) \\
&\hspace{1in} \times \left( \tilde D_\rho X^c + i \tilde D_\rho X^{c+4} \right),
\end{split}
\end{align}
and
\begin{align}
&\phantomequals \frac{36}{5} \frac{\pi \mu_2 \lambda^2}{k} C^{(3) \dagger}_{\mu A B} C^{(1)}_{\nu C \bar D} \beta^{AB \dagger}_C D_\rho Y^D \\
&= \frac{36}{5} \frac{\pi \mu_2 \lambda^2}{k}  C^{(3) \dagger}_{\mu a b} C^{(1)}_{\nu 4 \bar c} \beta^{a b \dagger}_4 D_\rho Y^c + 2 C^{(3) \dagger}_{\mu a 4} C^{(1)}_{\nu b \bar c} \beta^{a 4 \dagger}_b D_\rho Y^c \\
\begin{split}
&\rightarrow \tfrac{9}{10} \mu_2 \tilde \lambda^2  \Big( - C^{(3) \dagger}_{\mu a b} C^{(1)}_{\nu 4 \bar c} \left( [X^a, X^b] - i[X^{a+4}, X^b] - i[X^a, X^{b+4}] - [X^{a+4}, X^{b+4}] \right) \\
&\hspace{1in} \times \left( \tilde D_\rho X^c + i \tilde D_\rho X^{c+4} \right) \\
& \hspace{0.8in} - 2 C^{(1)}_{\mu b \bar c} C^{(3) \dagger}_{\nu a 4}  \left( [X^a, X^b] - i[X^{a+4}, X^b] + i[X^a, X^{b+4}] + [X^{a+4}, X^{b+4}] \right) \\
&\hspace{1in} \times \left( \tilde D_\rho X^c + i \tilde D_\rho X^{c+4} \right)
\Big),
\end{split}
\end{align}
and
\begin{align}
&\phantomequals \frac{36}{5} \frac{\pi \mu_2 \lambda^2}{k}  C^{(3) \dagger}_{\mu A B} C^{(3)}_{\nu C D} \beta^{CD}_A D_\rho Y^\dagger_B \\
&= \frac{36}{5} \frac{\pi \mu_2 \lambda^2}{k} \left( 2 C^{(3) \dagger}_{\mu b c} C^{(3)}_{\nu a 4} \beta^{a 4}_b D_\rho Y^\dagger_c + C^{(3) \dagger}_{\mu 4 c} C^{(3)}_{\nu a b} \beta^{a b}_4 D_\rho Y^\dagger_c  \right) \\
\begin{split}
&\rightarrow \tfrac{9}{10} \mu_2 \tilde \lambda^2 \Big(
2 C^{(3) \dagger}_{\mu b c} C^{(3)}_{\nu a 4} \left( [X^a, X^b] + i[X^{a+4}, X^b] - i[X^a, X^{b+4}] + [X^{a+4}, X^{b+4}] \right) \\
&\hspace{1in} \times \left( \tilde D_\rho X^c + i \tilde D_\rho X^{c+4} \right) \\
&\hspace{0.8in} + C^{(3)}_{\mu a b} C^{(3) \dagger}_{\nu 4 c}  \left( [X^a, X^b] + i[X^{a+4}, X^b] + i[X^a, X^{b+4}] - [X^{a+4}, X^{b+4}] \right)  \\
&\hspace{1in} \times \left( \tilde D_\rho X^c + i \tilde D_\rho X^{c+4} \right)
\Big).
\end{split}
\end{align}
These can all be identified with the known terms in the D$2$ action given in equation \eqref{eq:d2 1 derivative mixed indices}.

\section{Discussion and Conclusions}

So far we have only identified couplings in the M$2$-brane action which recover a small number of the terms making up the full $\tilde C \wedge \tilde B$ part of the D$2$ brane action. We have fully matched the terms with no derivatives and considered enough terms to motivate the one derivative terms. Combining all of the terms considered so far, we have identified the following additional couplings at quadratic order in the background fields:

\begin{align}
\begin{split}
S &= \frac{6}{5} \frac{\pi \mu_2 \lambda}{k} \int \varepsilon^{\mu \nu \rho} \Tr \bigg( 
- C_{\mu \nu \rho} C^{(1)}_{A B C} \gamma^{ABC} + C_{\mu \nu \rho} C^{(2)}_{A B \bar{C}} \beta^{AB}_{C} - C_{\mu \nu \rho} C^{(1) \dagger}_{A B C} \gamma^{ABC \dagger} + C_{\mu \nu \rho} C^{(2) \dagger}_{A B \bar{C}} \beta^{AB \dagger}_{C} \\
&\qquad + 3 \left( 2C_{\mu \nu A} C^{(1)}_{\rho B \bar{C}} \beta^{AB}_C + 3 C_{\mu \nu A} C^{(3)}_{\rho B C} \gamma^{ABC} - 2 C_{\mu \nu A} C^{(1) \dagger}_{\rho C \bar{B}} \beta^{AB}_C + C_{\mu \nu A} C^{(3) \dagger}_{\rho B C} \beta^{B C \dagger}_A \right) \\
&\qquad + 3 \lambda \bigg( 3C_{\mu \nu A} C^{(1)}_{B C D} \gamma^{ABC} D_{\rho} Y^D - C_{\mu \nu D} C^{(1)}_{A B C} \gamma^{ABC} D_{\rho} Y^D + 3 C_{\mu \nu A} C^{(2)}_{B C \bar D} \gamma^{ABC} D_{\rho} Y^\dagger_D \\
&\qquad \qquad + 2 C_{\mu \nu A} C^{(2)}_{B C \bar D} \beta^{AB}_{D} D_{\rho} Y^C + C_{\mu \nu C} C^{(2)}_{A B \bar D} \beta^{AB}_{D} D_{\rho} Y^C \bigg) \\
&\qquad + 6 \lambda \bigg( - 2 C^{(3)}_{\mu A B}C^{(1)}_{\nu C \bar D} \beta^{BC}_D D_\rho Y^A - C^{(3)}_{\mu B C}C^{(1)}_{\nu A \bar D} \beta^{BC}_D D_\rho Y^A - 3 C^{(3)}_{\mu A B} C^{(1)}_{\nu C \bar D} \gamma^{ABC} D_\rho Y^\dagger_D \\
&\qquad \qquad - 2 C^{(1)}_{\mu A \bar B} C^{(1)}_{\nu C \bar D} \beta^{A C}_B D_\rho Y^{\dagger}_D - 2 C^{(1)}_{\mu A \bar B} C^{(1)}_{\nu C \bar D} \beta^{B D \dagger}_A D_\rho Y^C \\
&\qquad \qquad + 3 C^{(3)}_{\mu A \bar B} C^{(3)}_{\nu C \bar D} \gamma^{A B C} D_\rho Y^D + 3 C^{(3) \dagger}_{\mu A B} C^{(1)}_{\nu C \bar D} \gamma^{ABD \dagger} D_\rho Y^C\\
&\qquad \qquad + C^{(3) \dagger}_{\mu A B} C^{(1)}_{\nu C \bar D} \beta^{AB \dagger}_C D_\rho Y^D + C^{(3) \dagger}_{\mu A B} C^{(3)}_{\nu C D} \beta^{CD}_A D_\rho Y^\dagger_B \bigg)
\bigg) 
\end{split}
\end{align}

Matching the two actions fully via the expand and compare method we have been using above is infeasible due to the growth in the number of terms with each additional derivative. However, there is evidence that it is possible to recover all of the terms in the $\tilde C \wedge \tilde B$ piece of the D$2$ action. The explicit identifications shown above have required a number of unexpected and non-trivial simplifications and we have qualitatively seen all of the features which will be involved in the full identification process. For example, terms which naively seem to appear in D$2$ action but have no way to be recovered from the M$2$ action have naturally cancelled within the D$2$ action itself. 
It is reassuring that the introduction of terms which mix daggered and undaggered $C$ fields was both natural in the M$2$ action and required to produce the D$2$ action. We also note that the parts of the D$2$ action given by $\tilde C$ and $\tilde B$ fields which both contain a sum of more than one M$2$ field and thus contain multiple cross terms have also had many terms checked explicitly. 
We have been free to choose the M$2$ coupling coefficients to obtain the correct factors in the D$2$ action. However, many of the terms that have been identified come from a mixture of more than one original term and we have no control over the required ratios between the coefficients. Fortunately the coefficients we have had to fix do come with the correct ratios to make this possible.

It seems there is some underlying principle guiding the construction and reduction of such an action and based on these observations we believe the identification will be possible to carry out for all quadratic couplings in the M$2$ brane, up to and including the $3$ derivative terms.

The introduction of an anti-symmetrised product between the $C$ fields is natural from the M$2$ point of view and is also required to fix the correct sign and order on many of the terms arising in the reduction. However, the recovered D$2$ action must then also have an anti-symmetric product between $\tilde C$ and $\tilde B$. For this to be consistent we need to consider $\tilde C \wedge \tilde B$ as a single object in the symmetrised trace on the D$2$-brane. Indeed this should be the case since we can produce or absorb the $\tilde C_{(3)} \wedge \tilde B$ by a suitable redefinition of $\tilde C_{(5)}$.

It is still an open question how to construct an extension to the ABJM action that preserves the full $U(N) \times U(N)$ gauge symmetry and reduces to the terms identified in this paper after the gauge group is broken to a single $U(N)$. 
An answer to this will hopefully also provide some indication of the form of the full coupling, without having to match every coefficient by hand with the D$2$ action.
Further understanding of how the action should behave under background gauge transformations may provide useful insights into this. The standard pull-back of $C$ involving partial derivatives will transform as a total derivative under the background gauge transformation $C \rightarrow C + d\Lambda$ and thus leave the action invariant. However, the pull-back involving covariant derivatives no longer transforms as an overall total derivative and we should instead view the pull-back as an object with its own transformation under background field transformations \cite{Adam:2003uq}. We can also view the pull-back as the procedure which uplifts the background fields to matrix valued fields on the multiple brane action. We need to understand better how this works for ABJM action with its scalars in the fundamental representation of $SU(4)$.

It is worth noting that the above procedure only recovers $\tilde C \wedge \tilde B$ terms but the D$2$ action also contains $\tilde C \wedge F$ terms. It is not clear how the novel way in which the Yang-Mills term appears in reduction can be extended to produce a term which couples the field strength to the background fields.

\section*{Acknowledgements}

We would like to thank Chong-Sun Chu and Gurdeep Sehmbi for useful discussions.
JPA is supported by an STFC studentship and the work of DJS is supported in part
by the STFC.

\appendix

\section{Identification of SU(4) fields with SO(7) fields} \label{identification of field components}

We present here a list of the identifications made in \cite{Kim:2010hj} between the components of the $SU(4)$ and $SO(7)$ fields. In this note we have scaled $\tilde B_{\mu \nu} \rightarrow 3 \tilde B_{\mu \nu}$ and $\tilde B_{\mu i} \rightarrow \frac{3}{2} \tilde B_{\mu i}$ compared to the original paper. This gives the pull back of $\tilde B$ a more natural form:
\begin{align}
P[\tilde B_{\mu\nu}]= \tilde B_{\mu\nu}+ 2 \lambda
\llangle \tilde B_{\mu i} \tilde D_\nu \tilde X^i \rrangle
+ \lambda^2 \llangle \tilde B_{ij} \tilde D_\nu \tilde X^i\tilde D_\rho \tilde X^j \rrangle.
\end{align}

The identifications used in this note are then:
\begin{align}
\tilde C_{\mu \nu \rho} &= C_{\mu \nu \rho} + C^{\dagger}_{\mu \nu \rho}, \\
\tilde B_{\mu \nu} &= 3 i \left( C_{\mu \nu 4} - C^\dagger_{\mu \nu 4} \right),
\end{align}
\begin{align}
\tilde C_{\mu \nu a} &= C_{\mu \nu a} + C^{\dagger}_{\mu \nu a}, \\
\tilde C_{\mu \nu a + 4} &= i \left( C_{\mu \nu a} - C^{\dagger}_{\mu \nu a}\right), \\
\tilde C_{\mu \nu 4} &= C_{\mu \nu 4} + C^{\dagger}_{\mu \nu 4}
\end{align}
\begin{align}
\tilde B_{\mu 4} &= 6i\big(C^{(1)}_{\mu4\bar4} - C^{(1)\dagger}_{\mu 4\bar 4}\big), \\
\tilde B_{\mu a} & = 3i\big(C^{(1)}_{\mu 4\bar a}-C^{(1)\dagger}_{\mu 4\bar
a}+ C^{(1)}_{\mu a\bar4}-C^{(1)\dagger}_{\mu a\bar 4}+C^{(3)}_{\mu 4a}
-C^{(3)\dagger}_{\mu 4a}\big), \\
\tilde B_{\mu a+4} &= 3\big(C^{(1)}_{\mu 4\bar a}+C^{(1)\dagger}_{\mu
4\bar a}- C^{(1)}_{\mu a\bar4}-C^{(1)\dagger}_{\mu a\bar 4}
-C^{(3)}_{\mu 4a} -C^{(3)\dagger}_{\mu 4a}\big), \\
\tilde C_{\mu4a} &= C^{(1)}_{\mu 4\bar a}
+C^{(1)\dagger}_{\mu 4\bar a}- C^{(1)}_{\mu a\bar4}-
C^{(1)\dagger}_{\mu a\bar 4} + C^{(3)}_{\mu 4a} + C^{(3)\dagger}_{\mu 4a}, \\
\tilde C_{\mu4a+4} &= -i\big(C^{(1)}_{\mu 4\bar a}
-C^{(1)\dagger}_{\mu 4\bar a} + C^{(1)}_{\mu a\bar4}
-C^{(1)\dagger}_{\mu a\bar 4} - C^{(3)}_{\mu 4a} + C^{(3)\dagger}_{\mu 4a}\big), \\
\tilde C_{\mu ab} &=
C^{(1)}_{\mu a \bar b} - C^{(1) \dagger}_{\mu b \bar a}
- C^{(1)}_{\mu b \bar a} + C^{(1)\dagger}_{\mu a \bar b}
+ C^{(3)\dagger}_{\mu a b} + C^{(3)}_{\mu a b}, \\
\tilde C_{\mu ab+4}
&=i\big(C^{(1)\dagger}_{\mu a\bar b}-C^{(1)}_{\mu b\bar a}
-C^{(1)}_{\mu a\bar b}+C^{(1)\dagger}_{\mu b \bar a}-C^{(3)\dagger}_{\mu ab}
+C^{(3)}_{\mu ab}\big), \\
\tilde C_{\mu a+4 b}
&=i\big(C^{(1)}_{\mu b\bar a} - C^{(1)\dagger}_{\mu a\bar b}
+C^{(1)}_{\mu a\bar b}  - C^{(1)\dagger}_{\mu b \bar a}
+ C^{(3)}_{\mu ab}\big) - C^{(3)\dagger}_{\mu ab} , \\
\tilde C_{\mu a+4 b+4} &=C^{(1)\dagger}_{\mu a\bar b}-C^{(1)}_{\mu
b\bar a}+ C^{(1)}_{\mu a\bar b}-C^{(1)\dagger}_{\mu b\bar a}
-C^{(3)\dagger}_{\mu a b} -C^{(3)}_{\mu ab}.
\end{align}
\begin{align}
\tilde B_{a4}&=-6i\big(C^{(2)}_{a4\bar 4}-C^{(2)\dagger}_{a4\bar 4}\big), \\
\tilde B_{4a+4}&= -6\big(C^{(2)}_{a4\bar 4}+C^{(2)\dagger}_{a4\bar 4}\big),\\
\tilde B_{ab} &= 3i \big(C^{(1)}_{ab4}-C^{(1)\dagger}_{ab4}
-C^{(2)}_{ab\bar 4}+ C^{(2)\dagger}_{ab\bar 4}
+C^{(2)}_{b4\bar a}- C^{(2)\dagger}_{ b4\bar a}
-C^{(2)}_{a4\bar b} +C^{(2)\dagger}_{a4\bar b}\big),\\
\tilde B_{ab+4} &= - 3\big(
C^{(1)}_{ab4}+C^{(1)\dagger}_{ab4}
-C^{(2)}_{ab\bar 4}- C^{(2)\dagger}_{ab\bar 4}
+C^{(2)}_{b4\bar a}+ C^{(2)\dagger}_{ b4\bar a}
+C^{(2)}_{a4\bar b} +C^{(2)\dagger}_{a4\bar b}\big), \\
\tilde B_{a+4 b} &= - 3\big(
C^{(1)}_{ab4}+C^{(1)\dagger}_{ab4}
-C^{(2)}_{ab\bar 4}- C^{(2)\dagger}_{ab\bar 4}
-C^{(2)}_{b4\bar a}- C^{(2)\dagger}_{ b4\bar a}
-C^{(2)}_{a4\bar b}- C^{(2)\dagger}_{a4\bar b} \big), \\
\tilde B_{a+4b+4} &= -3i \big(C^{(1)}_{ab4}-C^{(1)\dagger}_{ab4}
-C^{(2)}_{ab\bar 4}+ C^{(2)\dagger}_{ab\bar 4}
-C^{(2)}_{b4\bar a}+ C^{(2)\dagger}_{ b4\bar a}
+C^{(2)}_{a4\bar b} -C^{(2)\dagger}_{a4\bar b}\big), \\
\tilde C_{ab4} &= \big(
C^{(1)}_{ab4}+C^{(1)\dagger}_{ab4}
+C^{(2)}_{ab\bar 4}+ C^{(2)\dagger}_{ab\bar 4}
+C^{(2)}_{b4\bar a}+ C^{(2)\dagger}_{ b4\bar a}
-C^{(2)}_{a4\bar b} -C^{(2)\dagger}_{a4\bar b}\big), \\
\tilde C_{a4b+4} &= -\big(
C^{(1)}_{ab4}-C^{(1)\dagger}_{ab4}
+C^{(2)}_{ab\bar 4}- C^{(2)\dagger}_{ab\bar 4}
+C^{(2)}_{b4\bar a}- C^{(2)\dagger}_{ b4\bar a}
+C^{(2)}_{a4\bar b} -C^{(2)\dagger}_{a4\bar b}\big), \\
\tilde C_{4a+4b+4} &= -\big(
C^{(1)}_{ab4}+C^{(1)\dagger}_{ab4}
+C^{(2)}_{ab\bar 4}+ C^{(2)\dagger}_{ab\bar 4}
-C^{(2)}_{b4\bar a}- C^{(2)\dagger}_{ b4\bar a}
+C^{(2)}_{a4\bar b} +C^{(2)\dagger}_{a4\bar b}\big), \\
\tilde C_{abc} &=
C^{(1)}_{abc}+C^{(1)\dagger}_{abc}
+C^{(2)}_{ab\bar c}+ C^{(2)\dagger}_{ab\bar c}
-C^{(2)}_{ac\bar b}- C^{(2)\dagger}_{ ac\bar b}
+C^{(2)}_{bc\bar a} +C^{(2)\dagger}_{bc\bar a}, \\
\tilde C_{abc+4} &=i\big(
C^{(1)}_{abc}- C^{(1)\dagger}_{abc}
-C^{(2)}_{ab\bar c}+ C^{(2)\dagger}_{ab\bar c}
-C^{(2)}_{ac\bar b}+ C^{(2)\dagger}_{ ac\bar b}
+C^{(2)}_{bc\bar a} -C^{(2)\dagger}_{bc\bar a}\big), \\
\tilde C_{ab+4c+4} &=-\big(
C^{(1)}_{abc}+ C^{(1)\dagger}_{abc}
-C^{(2)}_{ab\bar c}- C^{(2)\dagger}_{ab\bar c}
+C^{(2)}_{ac\bar b}+ C^{(2)\dagger}_{ ac\bar b}
+C^{(2)}_{bc\bar a} +C^{(2)\dagger}_{bc\bar a}\big), \\
\tilde C_{a+4b+4c+4} &=-i\big(
C^{(1)}_{abc}- C^{(1)\dagger}_{abc}
-C^{(2)}_{ab\bar c}+ C^{(2)\dagger}_{ab\bar c}
+C^{(2)}_{ac\bar b}- C^{(2)\dagger}_{ ac\bar b}
-C^{(2)}_{bc\bar a} +C^{(2)\dagger}_{bc\bar a}\big).
\end{align}

Note that there is also a factor of two difference in the expressions for $\tilde B_{a4}$ and $\tilde B_{a+4 4}$ and a difference of sign of the $C^{(3)\dagger}$ parts of $C_{\mu i j}$. We believe these are mistakes in the original paper.

\bibliographystyle{utphys}
\bibliography{bibliography}

\end{document}